\documentclass[12pt,a4paper]{article}
\usepackage{empheq}
\usepackage{amsmath,amscd}
\usepackage{mathrsfs}
\usepackage{empheq}
\usepackage{amsfonts}
\usepackage{graphicx}
\usepackage{amssymb,scalerel,stackengine}
\usepackage[utf8]{inputenc}
\usepackage{color}
\usepackage{parskip}
\usepackage[table]{xcolor}
\usepackage{float}
\usepackage{enumitem}
\setlist[description]{font=\normalfont\itshape}
\usepackage[most]{tcolorbox}
\newtcbox{\othermathbox}[1][]{nobeforeafter, math upper, tcbox raise base, 
	enhanced, rounded corners, colback=black!5, colframe=black}
\usepackage{authblk}   
\usepackage[margin=1in]{geometry}
\usepackage{graphicx}
\usepackage[leftcaption]{sidecap}
\usepackage[font=small,labelfont=bf]{caption}
\usepackage{subcaption}
\usepackage{here}
\usepackage[colorlinks=true,allcolors=blue]{hyperref}
\usepackage{cite}
\usepackage{ragged2e}
\usepackage{etoolbox}

\apptocmd{\thebibliography}{\justifying}{}{}

\usepackage{tikz}
\usetikzlibrary{arrows.meta,calc,decorations.markings}

\let\OLDthebibliography\thebibliography
\renewcommand\thebibliography[1]{
	\OLDthebibliography{#1}
	\setlength{\parskip}{0pt}
	\setlength{\itemsep}{4.85pt plus 0.3ex}
}

\newcommand\beq{\begin{equation}}
	\newcommand\ee{\end{equation}}
\newcommand\cO{{\cal O}}

\newcommand\ie{\textit{i.e.}\ }
\newcommand\eg{\textit{e.g.}\ }

\def\hk{{\hat k}}

\def\mn{{\mu\nu}}

\def\de{\delta}

\def\pd{\partial}

\def\cd{\nabla}
\def\eps{\epsilon}

\def\cC{\mathcal{C}}

\def\cF{\mathcal{F}}

\def\cP{\mathcal{P}}

\def\w{{\omega}}

\def\bw{\boldsymbol{\omega}}

\def\cO{\mathcal{O}}

\def\bB{\boldsymbol{B}}

\def\bE{\boldsymbol{E}}

\def\bS{\boldsymbol{S}}

\def\re{\mathcal{R}e}
\def\th{\theta}

\def\lgw{\lambda_{\rm g}}
\def\lem{\lambda_{\rm e}}
\def\tlg{\tilde{g}}
\def\cA{\mathcal{A}}
\def\cB{\mathcal{B}}
\def\cE{\mathcal{E}}
\def\cF{\mathcal{F}}

\newcommand*\xbar[1]{%
	\hbox{%
		\vbox{%
			\hrule height 0.5pt 
			\kern0.5ex
			\hbox{%
				\kern-0.1em
				\ensuremath{#1}%
				\kern-0.1em
			}%
		}%
	}%
}

\title{\bf Gravitational Holonomy in \\ Sagnac Interferometry}

\author[1]{Reza Javadinezhad}
\author[2,3]{Ali Seraj}

\affil[1]{\small\textit{School of Physics, Institute for Research in Fundamental Sciences (IPM),\newline P.O.Box 19395-5531, Tehran, Iran}}
\affil[2]{\textit{School of Quantum Physics and Matter, Institute for Research in Fundamental Sciences\newline  (IPM),P.O.Box 19395-5531, Tehran, Iran}}
\affil[3]{\textit{Leuven Gravity Institute, KU Leuven, Celestijnenlaan 200D box 2415, 3001 Leuven, Belgium}\\
\texttt{\href{ali\_seraj@ipm.ir}{ali\_seraj@ipm.ir}, \href{javadinezhad@ipm.ir}{javadinezhad@ipm.ir}}}

\date{}

\begin{document}
	
	\maketitle
	
	\begin{abstract}
		We analyze the influence of gravitational waves on a Sagnac interferometer formed by the interference of two counter-propagating beams traversing a closed spatial loop. In addition to the well-known Sagnac phase shift, we identify an additional contribution originating from a relative rotation in the polarization vectors. We formulate this effect as a gravitational holonomy associated to the internal Lorentz group. The magnitude of both effects is computed due to gravitational waves generated by a localized source far from the detector, at leading order in the inverse distance. For freely falling observers, the phase shift is zero and the polarization rotation becomes the dominant effect. 
	\end{abstract}
	
	\tableofcontents
	
	\section{Introduction}
	Consider two light beams propagating along the same spatial path in opposite directions. Assuming that the two beams are initially in phase, one could measure possible differences in the final phases, after completing one or several loops, by interfering the beams. In 1913, Sagnac proposed this setup and identified a phase difference if the setup rotates with a nonzero angular velocity\cite{Sagnac1913a,Sagnac1913b}. Early developments on the \textit{Sagnac effect} can be found in~\cite{Post1967}. The effect is  present in a large variety of experimental setups from ring--laser gyroscopes to matter--wave
	interferometers~\cite{Stedman1997,Anderson1994,RizziRuggiero2003a,RizziRuggiero2003b}. Being sensitive to rotations, Sagnac interferometer is used as an \textit{optical gyroscope} and is widely used in navigation systems.

	In general relativity, rotating bodies (such as a black hole or a  star) induce a time delay between counter-propagating light beams through the off-diagonal time–space components of the metric. This gravitational Sagnac effect provides an operational definition of rotation in general relativity and may be viewed as the gravitational analog of the electromagnetic (EM) Aharonov–Bohm effect~\cite{AshtekarMagnon1975,Anandan1981,cohen1993standard}. It is one manifestation of the broader class of gravitational effects influencing the propagation of EM waves, including the gravitational Faraday and spin-Hall effects~\cite{Plebanski1960,Ishihara1988,Sereno2004,Sereno2005,Faraoni2008,Lyutikov2017,Shoom:2020zhr,Shoom:2022oer}. The former corresponds to a rotation of the light’s polarization, while the latter induces a helicity-dependent modification of the light’s trajectory.
	The gravitational Sagnac effect due to the rotation of the Earth has been observed in GPS systems~\cite{allan1985around}. Moreover, in the context of gravitational lensing~\cite{Schneider:1992bmb}, the rotation of the lens produces a spin-dependent time delay between co- and counter-rotating light paths, which can be understood as a gravitational Sagnac effect~\cite{Ruggiero:2023ker}.
	
	In a different front, it is well-known that gravitational wave (GW) can leave a persistent imprint on spacetime as well as probe physical systems,
	known as gravitational memory effects~\cite{ZeldovichPolnarev1974,Christodoulou1991,Blanchet:1992br,Thorne1992,WisemanWill1991,Favata2009a,Favata2009b,Favata2010,Flanagan:2018yzh}. The well-known example is the permanent displacement between freely falling test masses; but several other memory effects have been discussed in the literature, after the discovery of the correspondence between memory effects, soft theorems and asymptotic symmetries~\cite{Strominger:2017zoo}. In particular, the spin memory effect~\cite{Pasterski:2015tva,Nichols:2017rqr}, and the gyroscopic memory~\cite{Seraj:2021rxd,Seraj:2022qyt,Seraj:2022qqj,Faye:2024utu} refer, respectively, to a time-delay between counter-orbiting light beams in the transverse plane of the GW, while the latter is a net change of orientation in a freely falling distant gyroscope. These two effects are closely related and connected to the gravitomagnetic effects of GWs.  One of our motivations is to construct a setup in which memory observables can be studied more systematically.
	
	Given its sensitivity to gravitomagnetic effects, the aim of the present paper is to carefully study the effect of GWs on the Sagnac interference pattern. It turns out that the application of Sagnac effect in GW detection has already been suggested in early works~\cite{Anandan1981}, and studied in great detail in later research~\cite{Ruggiero:2015gha,Sun:1996bj,Eberle:2010zz,Chen:2002mf,Tinto:2014lxa,Sivasubramanian2003,Frauendiener:2018gkw,Frauendiener:2018tut}. Yet, we will show that, in addition to the well-known relative time delay, there is an extra effect: GWs induce an asymmetric rotation in the polarization of counter-rotating beams, which affects the interference pattern. While the phase shift due to time delay is proportional to the frequncy of light, the latter effect is frequency-independent. In typical situations where the GW frequency is much smaller than the light's frequency, the polarization rotation is subdominant. Still,  it can be of theoretical importance, since it naturally encodes a gauge-invariant  holonomy of the internal Lorentz group ~\cite{Flanagan:2018yzh,Seraj:2022qqj}. 
	
	As a proof of principle, we compute both effects in the context of asymptotically flat spacetimes in Bondi-Sachs formalism to $O(r^{-1})$, \ie to leading order  in the distance to the source of GWs. In particular, we show that for a freely falling observer, the time-delay effect effect vanishes and the polarization rotation becomes the dominant observable Sagnac effect.

	\paragraph{Notation.} We work in natural units where $G=c=1$. We will use different indices listed in the Table~\ref{table 1}. 
	\begin{table}[h]
		\centering
		\begin{tabular}{|c|c|c|c|}
			\hline
			&  Spacetime& Local frame& range\\ \hline
			4d&  $\mu$,$\nu$, $\dots$& $a,b,c,\dots$ & $0-3$\\ \hline
			3d&  -& $i,j,k,\dots$ & $1-3$\\ \hline
			2d&  $A,B,\dots$& $I,J,K,\dots$& $2,3$ \\ \hline
		\end{tabular}
		\caption{Index notation  used in this paper.}
		\label{table 1}
	\end{table}
	
\begin{figure}[h]
		\centering
		\includegraphics[width=0.5\linewidth]{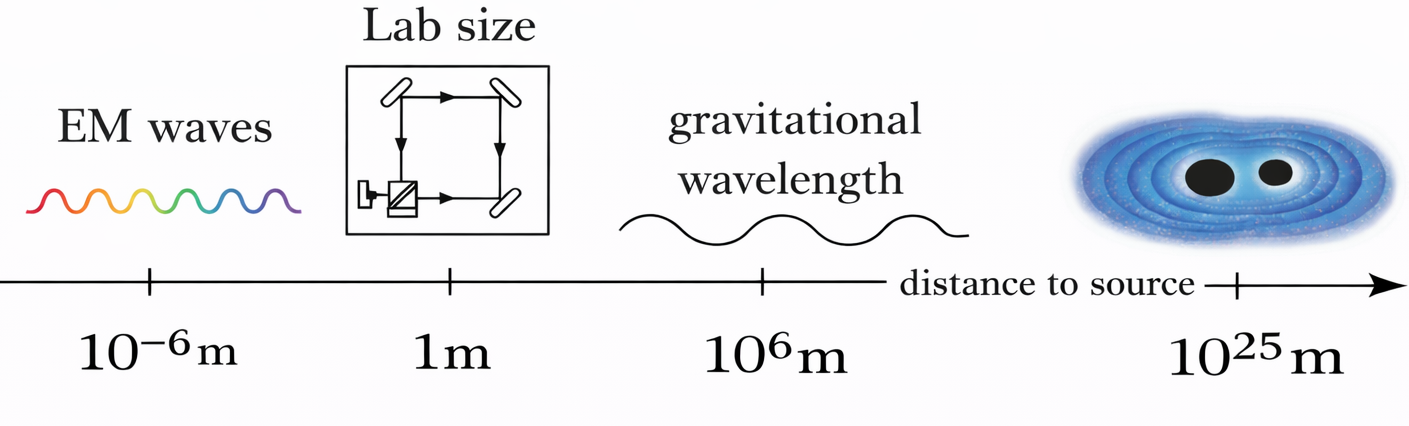}
		\caption{Vast hierarchy of length scales appearing in this problem.}
		\label{fig:hierarchy}
	\end{figure}

	Spatial vectors are denoted with boldface symbols, \eg $\bE$, with components  $E^i$. On the other hand, differential forms are typically not bold-faced, except a matrix-valued form defined in the text.
	\paragraph{Separation of scales.} 
	In this paper, we discuss the effect of gravitational waves on the propagation of EM waves, and in particular on the Sagnac interferometer. The problem is \textit{multiscale} in nature. With respect to the size $|X|$ of the interferometer, we have two regimes: the \textit{microphysics}, involving the EM wavelength $\lem$ (corresponding to the angular frequency $\w=\frac{2\pi}{\lem}$), and the \textit{macrophysics} involving the and the wavelength $\lgw$ of GWs and the distance $r$ between the observer and the source of GWs, with the following hierarchy which is depicted in Figure~\ref{fig:hierarchy},
	\begin{align}
		\lem\ll |X|\ll\lgw\ll r
	\end{align}
	
	To deal with this multiscale problem, we first use $\lem\ll |X|$ to formulate Maxwell equations in the eikonal approximation in Section~\ref{sec:eikonal}. This leads to the ray optics description of EM fields. Then, we use the hierarchy $|X|\ll\lgw\ll r$ to formulate the problem in the Fermi normal coordinates in Section \ref{sec: FNC} and apply it to the Sagnac experiment for distant observers in Section~\ref{sec: Bondi}.

	\section{Eikonal approximation and interference}\label{sec:eikonal}
	
	The \emph{eikonal} approximation provides a systematic framework for describing wave propagation, when the wavelength is much shorter than all other characteristic length scales in the problem. It is an important example of multiple-scale perturbation~\cite{Hinch_1991}. This approximation is widely used across diverse physical contexts, ranging from laser-beam propagation in fiber optics to gravitational lensing by compact astrophysical objects (see, \eg\cite{Schneider:1992bmb} for a textbook treatment). The central idea is that the waveform can be decomposed into a rapidly varying phase and a slowly varying amplitude. In the present work, we apply this framework to the propagation of EM waves on a dynamical curved spacetime that itself contains gravitational radiation. 
	
	Consider EM waves, described by a gauge field $A_\mu(x)$, propagating in a curved spacetime with metric $g_{\mn}(x)$. We expand the gauge field, in the eikonal form, as 
	\begin{align}\label{eikonal}
		A_\mu(x) \;=\; \re \left[a(x)\, f_\mu(x)\, e^{i S(x)/\epsilon}\right]\,,
	\end{align}
	where $\eps\ll 1$ is a bookkeeping parameter, $S(x)$ is the rapidly varying eikonal phase, $a(x)$ is the slowly varying scalar amplitude, and $f_\mu(x)$ is the slowly varying polarization vector with unit norm $f_\mu f^\mu = 1$. 
	We then insert this ansatz into the Maxwell's equations and solve the equations order by order in $\eps$. Define the wavevector $k_\mu \equiv \nabla_\mu S$ as the normal to the wavefronts. At $O(\eps^{-2})$, Maxwell equations reduce to the so-called \emph{eikonal equation}
	\begin{align}\label{eikonal equation}
		k^\mu k_\mu = 0\,,\qquad k_\mu = \nabla_\mu S\,.
	\end{align}
	These two properties imply that $k^\mu$ is geodesic as well
	\begin{align}
		k^\nu \nabla_\nu k_\mu = k^\nu \nabla_\mu k_\nu =\frac{1}{2}\cd_\mu (k^\mu k_\mu)=0.
	\end{align}
	Therefore, the EM waves propagate along null geodesics of the spacetime. 
	At next order $O(\eps^{-1})$, one finds transport equations for the polarization and the amplitude
	\begin{align}
		k^\nu \nabla_\nu f_\mu &= 0,\label{parallel transport}\qquad
		\nabla_\mu (a^2 k^\mu) = 0\,.
	\end{align}
	The former states that the polarization is parallel transported along the ray, while the latter is the conservation of photon number: the amplitude decreases when neighboring rays diverge, and increases when they converge.
	
	\paragraph{Transverse polarization.} The gauge field has a gauge redundancy, which can be partly fixed by imposing the Lorenz gauge condition $\cd^\mu A_\mu=0$, which implies, through the eikonal expansion, that $k^\mu f_\mu=0$. However, this does not completely fix the gauge redundancy. The Lorenz gauge is preserved by  gauge transformations $A_\mu\to A_\mu +\cd_\mu \lambda$ with $\cd_\mu \cd^\mu \lambda=0$. The latter is solved by $\lambda= \alpha(x) e^{i S(x)/\eps}$, where $\alpha(x)$ is arbitrary. Under this residual gauge symmetry, the polarization vector $f_\mu$ is transformed as
	\begin{align}\label{residual Lorenz}
		f_\mu\to f_\mu+\beta(x) k_\mu\,\qquad \beta=\frac{i\alpha}{a}\,.
	\end{align}
	The physical polarization can thus be  defined as the equivalence class $[f_\mu]$ with the equivalence relation $f_\mu\sim f_\mu+\beta  k_\mu$. The two conditions $k^\mu f_\mu=0$ and $f_\mu\sim f_\mu+\beta k_\mu$ implied by the gauge constraint leaves only two degrees of freedom in the polarization vector\footnote{Gauge symmetry, as a first-class constraint corresponds to a null hypersurface in the symplectic geometry of the phase space. Imposing the constraint kills one degree of freedom. Yet, the constraint generates gauge orbits tangent to the constraint submanifold, thereby reducing another freedom; see the elegant discussion in §2 of \cite{Henneaux:1994lbw}.}.
	
	Rather than working with equivalence classes,  we choose to perform a gauge fixing. To this end, we define an auxilliary null vector  $n^\mu$ such that $k\cdot n=-1$.  The purely transverse polarization vector $p^\mu$ is then defined as 
	\begin{align}
		p_\mu\equiv f_\mu +\beta k_\mu\,, \quad \text{such that} \quad p_\mu n^\mu=0  \,.
	\end{align}
	Imposing the latter condition, we find $\beta=f\cdot n$ and hence~\cite{Kulish:1970ut}
	\begin{align}\label{projector}
		p_\mu=\perp_\mu^{\;\;\nu} f_\nu\,,\qquad \perp_\mu^{\;\;\nu} =(\de_\mu{}^\nu+k_\mu n^\nu)\,. 
	\end{align}
	This equation identifies the physical polarization as the  transverse \textit{projection} of the naive polarization vector $f_\mu$, thanks to the properties 
	\begin{align}\label{projection properties}
		\perp_\mu^{\;\;\nu}k_\nu=0\,,\qquad \perp_\mu^{\;\;\alpha}\perp_\alpha^{\;\;\nu}=\perp_\mu^{\;\;\nu}\,.    
	\end{align}
	Crucially, the norm of the transverse polarization $p$ is identical to the norm of the parallel transported vector\footnote{Our result coincides with the massless limit of the definition of observable spin for massive particles in \cite{Damour:2007nc,Bini:2017xzy}. In the massive case, the canonical spin vector is obtained by a boost acting on the parallel transported spin four-vector. The boost is defined as the one taking the particle's velocity $u$ to that of the observer $\hat{t}$, while preserving the orthogonal 2-plane. The result is given in eqs. (14)-(17) of \cite{Bini:2017xzy}. Writing $u=\gamma (\hat{t}+v\hk)$, we observe that in the null limit $v\to 1$, their eq.(17) reduces to our \eqref{projector}.},
	\begin{align}\label{p norm}
		p_\mu p^\mu=f_\mu f^\mu=1\,.
	\end{align}
	For a given observer with normalized time vector field $\hat{t}^\mu$ (with $\hat{t}^\mu\hat{t}_\mu=-1$), there is always a preferred choice for the auxiliary vector $n$ as 
	\begin{align}\label{k,n decomposition}
		k=\w(\hat{t}+\hat{k})\,,\qquad n=\frac{1}{2\w}(\hat{t}-\hat{k})\,,
	\end{align}
	where $\hat{k}^\mu$ is the spatial direction of the propagating light ray (with $\hat{k}^\mu\hat{k}_\mu=1\,,\;\hat{t}^\mu\hat{k}_\mu=0$). This shows that $k^\mu,n^\mu$ trace the same spatial path but with opposite directions. Therefore, $n$ is proportional to the wavevector of the counter-propagating light ray. Given the decomposition \eqref{k,n decomposition}, the projector $\perp$ in \eqref{projector} is uniquely fixed by the tangent vector $\hk$. Another useful property of the projector is that under changing the propagation direction $\hat{k}\to -\hat{k}$, 
	\begin{align}\label{perp transpose}
		k\to 2\w^2 n\,,\qquad n\to \frac{1}{2\w^2}k\quad \implies\quad \perp_{-\hk}=\perp^\text{T}_{\hk}\,.
	\end{align}
	
	Suppose the observer is equipped with a comoving orthonormal frame 
	\begin{align}\label{observer frame}
		e_a{}^\mu=(e_0{}^\mu=\hat{t}^\mu , e_i{}^\mu) \,,\qquad \hat{t}\cdot e_i=0,\quad e_i\cdot e_j=\de_{ij}.
	\end{align}
	A vector $X$ can be decomposed, in the observer's frame, as 
	\begin{align}
		X=X^a e_a=X^0\, \hat{t}+X^i e_i\,,\qquad X^0\equiv -\hat{t}\cdot X,\quad X_i\equiv X\cdot e_i   \,.
	\end{align}
	One can decompose the wavevector  as $k=\w(\hat{t}+\hk^i e_i)$, or in components in the local frame  as $k^a=\w(1,\hk^i)$ and similarly $n^a=\frac{1}{2\w}(1,-\hat{k}^i)$. As a result, Eq.  \eqref{projector} can be rewritten in the local frame as
	\begin{align}\label{polarization spatial}
		p=p^i e_i\,,\qquad p_i=(\de_i{}^j-\hat{k}_i\hat{k}^j)f_j\,,
	\end{align}
	which is  the projection of $f^i$ on the plane transverse to the spatial propagation direction $\hk^i$. We will see shortly that  $p$ is the physical polarization vector that appears naturally in gauge invariant quantities such as the electric field and the Poynting vector.

	\subsection{Interference}\label{sec: interference}

	In any interferometric experiment, an EM beam is split into two beams through a beam splitter. The beams  travel along different spacetimes paths, say $\gamma_1,\gamma_2$. Finally the beams recombine (typically by the beam splitter itself) on a single path towards the detector's screen. 
	What is measured by the detector is the total light intensity $I_{\text{tot}}$ averaged over several cycles of the EM wave oscillation,
	\begin{align}
		I_{\text{tot}}&=\langle \bS_{\text{tot}}\cdot N \rangle\,,
	\end{align}
	where $\bS=\bE\times\bB$,  is the Poynting vector (not to be confused with the eikonal phase $S$), projected on the screen with normal vector $N$. 
	
	To describe each beam, we use the eikonal approximation \eqref{eikonal} implying
	\begin{align}
		F_\mn&=\re \left(\frac{i}{\eps}a e^{iS/\eps}(k_\mu f_\nu-k_\nu f_\mu)\right)+\cO(\eps^0)\,.
	\end{align}
	Imposing the Lorenz gauge $k^\mu f_\mu=0$ implies $f_0=-\hk^i f_i$, which can be used to write the electric field $E_i=F_{0i}=\hat{t}^\mu e_i{}^\nu F_\mn$ and magnetic field $B_i=\frac12\eps_{ijk} F_{jk}=\frac12\hat{t}^\mu e_i{}^\nu\eps_{\mn\alpha\beta}F^{\alpha\beta}$ as
	\begin{align}
		E_i= \re \left(\frac{a\omega}{i\eps} e^{iS/\eps}p_i\right)+\cO(\eps^0)\,,\qquad \bB=\hat{k}\times \bE+\cO(\eps^0)\, ,
	\end{align}
	where $p_i= (\de_i{}^j-\hk_i \hk^j)f_j$ is the transverse physical polarization that we introduced before in \eqref{projector} and \eqref{polarization spatial}.
	Accordingly, the Poynting vector reads $\bS=\bE\times (\hk\times \bE)+\cO(\eps^0)=|\bE|^2 \,\hk+\cO(\eps^0).$ 
	
	The two beams recombine towards the detector's screen, which is normal to the superposed beam. The total intensity is therefore
	\begin{align}\label{iint}
		I_{\text{tot}} &= \, \langle(\bE_1^2+\bE_2^2+2 \bE_1 \cdot \bE_2) \rangle=I_1+I_2+I_{int},
	\end{align}
	where $I=\langle\bE^2\rangle$ for each beam. The first two terms are the intensities of each of the beams. They can be measured independently using auxiliary monitoring detectors placed right before the two beams recombine at the beam splitter. The interference is then encoded in the \textit{cross-correlation function} $\gamma_{12}$ defined as 
	\begin{align}
		\gamma_{12}&\equiv\frac{(I_{\text{tot}} -I_1-I_2)}{2\sqrt{I_1\,I_2}}=\frac{\langle\bE_1 \cdot \bE_2 \rangle}{\sqrt{\langle |\bE_1|^2\rangle\langle |\bE_2|^2\rangle}}\,.
	\end{align}  
	We assume that the beams have linear polarization when interfered, implying that the polarization for the two beams are real valued. Accordingly, for each beam,
	\begin{align}\label{E,B}
		E_i&=-\frac{a\omega}{\eps} \sin (S/\eps)\, p_i\,,\qquad \sqrt{2I}=\frac{a\omega}{\eps}\,,
	\end{align}
	where in the last equality, we used $|p|=1$, thanks to \eqref{p norm}.
	As a result, the cross-correlation function becomes
	\begin{empheq}[box=\othermathbox]{equation}\label{cross-correlation}
		\gamma_{12}=\cos\Delta S \,\cos\psi \,,\quad \Delta S\equiv S_1-S_2\,,\quad  \cos\psi\equiv p_1\cdot p_2\,.
	\end{empheq}
	Therefore, the interferometric effect is purely written in terms of two angles: I) the difference $ \Delta S$ in the eikonal phases of the two beams , II) the angle $\psi =\cos ^{-1}(p_1\cdot p_2)$ between the polarization of the beams when they interfere.  
	Note the disappearance of $\eps$ in \eqref{cross-correlation}. This is because while each of $S_1, S_2$ are fast variables, the difference in phases $S_1-S_2$ is a slow/adiabatic effect proportional to $\eps$ which is cancelled by the $\eps^{-1}$ in the eikonal exponent.

	\paragraph{Initial conditions.} The interference described by \eqref{cross-correlation} crucially depends on the initial phases and polarizations of the two beams. If the two beams are initially in phase and have identical polarization vectors, then both angles $\Delta S,\psi$ are initially zero and get a small correction after completing the loop. As a result,
	\begin{align*}
		\gamma_{12}&\approx 1-\frac{1}{2}\big(\Delta S^2+\psi^2\big)\,.
	\end{align*}
	Being quadratic in the small angles, the interferometric effect is highly suppressed and hence hard to detect. This suppression due to cosine function can be circumvented by changing the initial states. One can introduce an initial relative phase or an initial rotation of polarization by using  a suitable optical device. In particular, it is wise to prepare the initial state of the beams to have either orthogonal phases or orthogonal polarization vectors. Let us discuss these separately.
	\begin{itemize}[leftmargin=0pt]
		\item \textit{Orthogonal phase/aligned polarization}
		\begin{align}
			\Delta S\big\vert_{t=0}=\pi/2, \qquad p_1\cdot p_2\big\vert_{t=0}=1,    
		\end{align}
		In this case, the setup is sensitive to phase difference as the polarization effect is second order, and
		\begin{align}
			\gamma_{12}&\approx \pi/2-\Delta S.
		\end{align}
		\item \textit{Aligned phase/Orthogonal polarization}
		\begin{align}
			\Delta S\big\vert_{t=0}=0, \qquad p_1\cdot p_2\big\vert_{t=0}=0   \, .
		\end{align}
		In this case, 
		\begin{align}\label{polarization sensitve}
			\gamma_{12}&=p_1\cdot p_2\approx \pi/2-\psi\, ,
		\end{align}
		where $\psi$ is the angle between the two polarization vectors. These are summarized in Table \ref{table:initial conditions}.
	\end{itemize}
	The above analysis shows that the initial conditions can be arranged such that the interferometer is sensitive to the relative polarization rotation, rather than the relative eikonal phase. We will show, in Section \ref{Sec:polarization}, that the latter measures the gravitational holonomy along a closed spacetime loop defined by the Sagnac setup. Accordingly, the Sagnac setup with suitable initial conditions can be considered as a \textit{detector of gravitational holonomy}.
	\begin{table}
		\centering
		\begin{tabular}{|c|c|c|c|}
			\hline
			\rowcolor{lightgray}  Initial phase  & Initial polarization  & cross-correlation& Sensitive to\\
			\hline
			$ \Delta S\vert_{t=0}=\pi/2$ & $ p _1\cdot p _2 \vert_{t=0}=1 $  & $\gamma_{12}\approx \pi/2-\Delta S$ & phase difference\\
			\hline
			$\Delta S\vert_{t=0} =0$ & $ p_1\cdot p_2\vert_{t=0} =0$  & $\gamma_{12}\approx \pi/2-\psi$ &polarization rotation\\
			\hline
		\end{tabular}
		\captionsetup{width=.8\linewidth}
		{\caption{Various choices of the initial states of the beams that magnify different physical quantities.\label{table:initial conditions}}}    
	\end{table}
	
	
	To conclude this section, let us stress on an observational distinction between the eikonal phase shift $\Delta S$ and the polarization rotation $\psi$. Decomposing the EM field in the circular polarization (helicity) basis, one can see that the two effects correspond to a phase shift in each polarization. The point is, however, that the eikonal phase shift $\Delta S$ rotates the two polarizations in the same way, while the phase induced by polarization rotation on right-handed photons is opposite that of left-handed ones. 
	This suggests an alternative way to empirically resolve the eikonal phase shift from the polarization rotation: one can split the light beams inside the Sagnac interferometer into its two polarization components and interfere them separately. Eq.  \eqref{cross-correlation}, can be written alternatively as 
	\begin{align}
		\gamma_{12}=\frac12\Big(\cos(\Delta S-\psi)+\cos(\Delta S+\psi)\Big)\,.
	\end{align}
	Each of the two terms on the right-hand side is actually the contribution from one of the polarizations.

	\subsection{Gravitational holonomy}\label{Sec:polarization}

	\paragraph{Evolution of the polarization.}
	It was shown in \eqref{parallel transport} that the polarization vector $f^\mu$ is parallel transported along the null ray. This equation can be solved in a local orthonormal frame $e_a{}^\mu$ and its inverse $e^a{}_\mu$ as follows.
	\begin{align}
		e_a{}^\mu\,:\qquad    g_{\mn}e_a{}^\mu e_b{}^\nu=\eta_{ab}\,,\qquad \eta_{ab}e^a{}_\mu e^b{}_\nu=g_\mn.
	\end{align}
	Write $f^\mu=f^a e_a^\mu$ and let the light trajectory be a worldline  $\gamma: \{x^\mu(\lambda)\}$, where $\lambda$ is a parameter on the null geodesic of the light ray.  The parallel transport equation can be recast for $f^a$ as 
	\begin{align}\label{fa evolution}
		\dfrac{df^a}{d\lambda}=\Omega^a{}_b f^b\,,\qquad \Omega^a{}_b=-k^\mu \omega_{\mu}{}^a{}_b,
	\end{align}
	where $\omega_{\mu}{}^a{}_b=e^a{}_\nu \cd_\mu e_b{}^\nu$ is the associated torsionless spin connection, which is real and antisymmetric under $a\leftrightarrow b$. The spin connection as a one-form is denoted as $\omega^a{}_b=\omega_{\mu}{}^a{}_b dx^\mu$ is , and  $\bw$ denotes the matrix valued one-form with elements $\omega^a{}_b\,$. Equation \eqref{fa evolution} is solved by a \textit{path-ordered exponential}
	\begin{align}\label{f evolution}
		f^a(t)={U}^a{}_{b}(t) \, f_0^b\,,\qquad U(t)=\cP \exp \left(-\int_0^t \bw\right).
	\end{align}
	where in integral is over the worldline $\gamma$ of the light ray and $f_0=f(\lambda=0)$ is the initial polarization. Note that the evolution matrix $U$ is path-dependent, and to stress this, we sometimes denote it as $U_\gamma$. Since $\bw$ is a one-form, it can be unambiguously integrated on the worldline. The Dyson series expansion can be used to approximate the path-ordered exponential
	\begin{align}
		U^a{}_b(t)=\de^a{}_b-\int_{0}^{t} \omega^a{}_b(\lambda) d\lambda\,+\int_0^{t}  \omega^a{}_c(\lambda)\int_0^{\lambda} \omega^c{}_b(\lambda') d\lambda d\lambda'+\cdots .
	\end{align}
	Holonomies have three important properties:
	\begin{description}[leftmargin=0pt]
		\item[\textit{Inversion.}] Since $\bw^\dagger=\bw^{\mathrm{T}}=-\bw$ is  antisymmetric,  $U_\gamma^\dagger=U_{-\gamma}=U_\gamma^{-1}$, where $U_{-\gamma}$ is the holonomy along the reversed path. 
		\item[\textit{Composition.}] Given two curves $\gamma_1$ and $\gamma_2$ such that the endpoint of $\gamma_1$ coincides with the starting point of $\gamma_2$, then
		\begin{align}
			U_{\gamma_2}U_{\gamma_1}=U_{\gamma_1\circ\gamma_2},
		\end{align}
		where $\gamma_1\circ\gamma_2$ is the joint curve of $\gamma_1$ and $\gamma_2$.
		\item[\textit{Gauge transformation.}] Under a local Lorentz transformation $\Lambda(x)\in SO(3,1)$ of the tetrad, $\boldsymbol{e}'=\Lambda \, \boldsymbol{e}$, which implies $\bw'=\Lambda \bw \Lambda^{-1} + \Lambda d(\Lambda^{-1})$. The holonomy $U_\gamma$ with endpoints $x_0,x_f$ then transforms nicely as 
		\begin{align}\label{holonomy gauge transformation}
			U'_{\gamma}=\Lambda(x_f)\, U_{\gamma} \,\Lambda^{-1}(x_0)\, .
		\end{align}
		These properties can be proven either by the definition of the holonomy in terms of the path-ordered exponential, or its corresponding transport equation $(\frac{d}{d\lambda}-\Omega) U=0$ with the initial condition $U=1$ at $\lambda=0$.
	\end{description}

	\section{Sagnac interferometer}\label{sec:Sagnac}
	
	\begin{figure}[ht]
		\centering
		\definecolor{colorone}{RGB}{31,122,140} 
\definecolor{colortwo}{RGB}{232,130,13} 

\tikzset{
	midarrow/.style={
		postaction={decorate},
		decoration={markings, mark=at position 0.5 with {\arrow{Stealth}}}
	},
	midarrow dashed/.style={
		dashed,
		postaction={decorate},
		decoration={markings, mark=at position 0.5 with {\arrow{Stealth}}}
	}
}

	\begin{tikzpicture}[
		line cap=round,
		line join=round,
		>=Stealth,
		font=\small
		,scale=0.85]
		
		\coordinate (BS)  at (0,0);    
		\coordinate (M2)  at (5,0);    
		\coordinate (M1)  at (5,3.5);  
		\coordinate (M3)  at (0,3.5);  
		\coordinate (LAS) at (-3,0);   
		\coordinate (DET) at (0,-1.5); 
		
		\def\bssize{0.9}
		
		\coordinate (BSL) at ($(BS)+(-\bssize/2,0)$);  
		\coordinate (BSR) at ($(BS)+(\bssize/2,0)$);   
		\coordinate (BSU) at ($(BS)+(0,\bssize/2)$);   
		\coordinate (BSD) at ($(BS)+(0,-\bssize/2)$);  
		
		\draw[thick,fill=gray!15] (BS) ++(-\bssize/2,-\bssize/2) rectangle ++(\bssize,\bssize);
		\draw[thick,fill=gray!15] (BS) ++(-\bssize/2,-\bssize/2) -- ++(\bssize,\bssize);
		
		\def\mirw{1.2}
		\def\mirh{0.25}
		
		\begin{scope}[shift={(M1)},shift={(0.1,0.1)},rotate=-45]
			\draw[thick,fill=gray!15] (-\mirw/2,-\mirh/2) rectangle ++(\mirw,\mirh);
		\end{scope}
		
		\begin{scope}[shift={(M2)},shift={(0.1,-0.1)},rotate=45]
			\draw[thick,fill=gray!15] (-\mirw/2,-\mirh/2) rectangle ++(\mirw,\mirh);
		\end{scope}
		
		\begin{scope}[shift={(M3)},shift={(-0.1,0.1)},rotate=45]
			\draw[thick,fill=gray!15] (-\mirw/2,-\mirh/2) rectangle ++(\mirw,\mirh);
		\end{scope}
		
		\def\dw{3}
		\def\dh{1}
		\draw[thick] ($(DET)+(-\dw/2,0)$) -- ($(DET)+(\dw/2,0)$);
		
		
		\draw[thick,midarrow] (LAS) -- (BSL)
		node[midway,below] {Laser};
		
		\draw[thick,midarrow,color=colorone] (BSR) -- (M2);
		\draw[thick,midarrow,color=colorone] (M2) -- (M1);
		\draw[thick,midarrow,color=colorone] (M1) -- (M3);
		\draw[thick,midarrow,color=colorone] (M3) -- (BSU);
		
		\draw[thick,midarrow,color=colortwo]
		($(BSU)+(0.18,0)$) --
		($(M3)+(0.18,0.18)$);
		\draw[thick,midarrow,color=colortwo]
		($(M3)+(0.18,0.18)$) --
		($(M1)+(-0.18,0.18)$);
		\draw[thick,midarrow,color=colortwo]
		($(M1)+(-0.18,0.18)$) --
		($(M2)+(-0.18,-0.18)$);
		\draw[thick,midarrow,color=colortwo]
		($(M2)+(-0.18,-0.18)$) --
		($(BSR)+(0,-0.18)$);
		
		\draw[thick,midarrow,color=colorone] (BSD) -- (DET);
		\draw[thick,midarrow,color=colortwo] ($(BSD)+(0.2,0)$)  -- ($(DET)+(0.2,0)$);
		
		\node[below left]  at (0.1,1.1) {BS};
		\node[below]       at ($(DET)+(0,-.1)$) {Screen};
		
	\end{tikzpicture}
	
		\captionsetup{width=.8\linewidth}
		\caption{Schematic figure of the  Sagnac interferometer. A laser beam enters the beam splitter (BS) and is split into two beams, which  travel in a closed spatial path in opposite directions. The path is guided by several mirrors. At the end of the trip, the recombine through the beam splitter and interfere on the detector's screen.}\label{fig:Sagnac}
	\end{figure}
	
\begin{figure}[h]
		\centering
		\begin{minipage}{0.2\textwidth}
			\includegraphics[width=\textwidth]{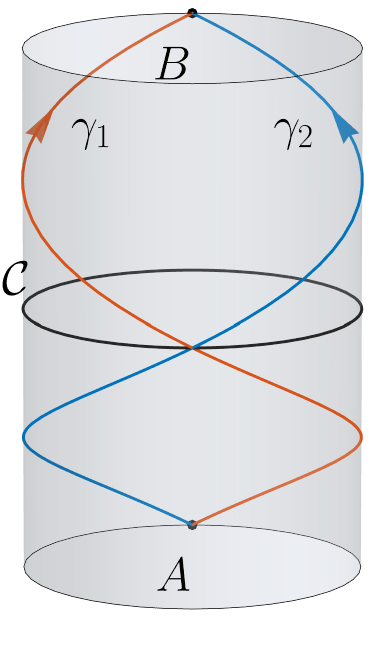}
		\end{minipage}\qquad
		\begin{minipage}{0.4\textwidth}
			\captionof{figure}{Two light rays propagate along null worldlines $\gamma_1,\gamma_2$, starting at spacetime event $A$, and recombine at event $B$. The joint curve $\Gamma=\gamma_1\circ -\gamma_2$ is the closed spacetime curve starting from A, going forward in time along $\gamma_1$ and returning back to $A$ along $\gamma_2$.\label{fig:holonomy}}
		\end{minipage}
	\end{figure}

	Let us now apply the general results above to a Sagnac interferometer. A light beam is divided by a beam splitter into two components that propagate along a common closed path $\cC$ in counter-propagating directions. Upon completing the circuit, the beams recombine (typically by the beam splitter itself) and exit the interferometer towards the detector's screen, resulting in an interference pattern. A schematic  of the setup is provided in Figure~\ref{fig:Sagnac}. The \textit{spacetime} picture is as follows: two light beams start from the same spacetime event, say $A$, trace different worldlines $\gamma_1,\gamma_2$ and recombine at another spacetime event $B$. This is depicted in Figure~\ref{fig:holonomy}. 
	
	\paragraph{Mirrors vs. fibers.} As we discussed in Section~\ref{sec:eikonal}, in vacuum, light propagates on null geodesics. In the weak gravity regime, which is the context of this paper, there is at most one null geodesic connecting two spacetime points $A,B$. Therefore, the Sagnac setup cannot be realized in vacuum. Instead, we will use \textit{piecewise} geodesic paths, guided by mirrors. This allows us to continue using the formalism of Section~\ref{sec:eikonal} in the interval between the reflections. One might worry how to include the effect of the reflection on the polarization. However, this issue can be also circumvented by using a suitable reflector for which there is no additional phase due to the reflection\footnote{For a glass/air reflection, the additional phase due to reflection is zero between the Brewster and the total reflection angles~\cite{Hecht2017}.}.
	Another option is to  guide light rays on non-geodesic paths, using a medium, like an optical fiber. In this case, one needs to generalize the construction of Section~\ref{sec:eikonal} to medium. We discuss this in Appendix \ref{appendix:fiber}.

	Each light beam is characterized by its tangent vector $k^\mu\equiv \frac{dx^\mu}{d\lambda}$, and its polarization $p^\mu$, given by \eqref{projector}. Expanding the projectors, one finds 
	\begin{align}
		p_2\cdot p_1&=(\perp_{\hat{k}_2}f_2)^T(\perp_{\hat{k}_1}f_1) = f_2^T\perp_{-\hat{k}_2}\perp_{\hat{k}_1}f_1=f_2^T\perp_{\hat{k}_1}f_1=
		f_1\cdot f_2+(f_1.n_1)(f_2.n_2),
	\end{align}
	where we have used the properties of the projection operator \eqref{projection properties} and \eqref{perp transpose}, together with the fact that the two beams propagate in opposite directions.
	Moreover, in the absence of gravitational effects one has \(f\!\cdot\! k = 0 = f\!\cdot\! n\), so all of the inner products appearing above are at least order \(O(r^{-1})\) and therefore conclude
	
	\begin{equation}\label{p1p2}
		p_2\cdot p_1=
		f_1\cdot f_2+O(r^{-2}).
	\end{equation}
	
	Since we are focused on leading effects, it is enough to compute $f_1\cdot f_2$, which we show in subsection \ref{section:Sagnac polarization} that it captures the gravitational holonomy.
	
	\subsection{Eikonal phase difference}\label{sec: eikonal phase sagnac}
	
	The eikonal equation $k^\mu k_{\mu}=g^{\mu\nu}\pd_\mu S \pd_\nu S=0$ can be solved perturbatively by using a metric perturbation around the flat metric,
	\begin{equation}
		g_{\mu\nu}=\eta_{\mu\nu}+\delta g_{\mu\nu}.
	\end{equation}
	Accordingly we write the eikonal phase as $S=S^{(0)}+S^{(1)}$ in the FNC. To leading order we find the equation for the flat space solution $S^{(0)}$,
	\begin{equation}
		\eta^{\mu\nu}\pd_\mu S^{(0)} \pd_\nu S^{(0)}=0.
	\end{equation}
	This equation admits a simple plane-wave solution of the form \(S^{(0)} = q_\mu X^\mu\).
	The four-vector \(q_\mu = (-\omega, q_i)\) is null in flat spacetime, satisfying
	\(q_\mu q_\nu \eta^{\mu\nu} = 0\).
	At first subleading order, the equation for \(S^{(1)}\) becomes,
	\begin{equation}\label{eikonal first order}
		q^\mu \pd_\mu S^{(1)}=-\frac{1}{2}q_\mu q_\nu \delta g^{\mu\nu}=\frac{1}{2}q^\mu q^\nu \delta g_{\mu\nu}=\frac{1}{2}(\omega^2\delta g_{00}+\delta g_{ij}\,q^i\,q^j)+\omega\, q^i \,g_{0i},
	\end{equation}
	where we have used the elemantry relation $\delta g_{\mu\nu}=-\eta_{\mu\alpha}\eta_{\nu\beta}\delta g^{\alpha\beta}$. What we ultimately wish to compute is the phase difference between two counter-propagating beams in the spacetime with inverse metric $g^{\mu\nu}$. 
	This phase difference vanishes for the leading eikonal \(S^{(0)}\) but is nonzero at subleading order \(S^{(1)}\).
	One may extract it by explicitly solving the subleading equation \eqref{eikonal first order}, but this is unnecessary, since the same information can be obtained by directly integrating that equation.
	Because the equation is already subleading, we may set \(q^\mu=\tfrac{dx^\mu}{d\lambda}\) and consistently neglect higher-order corrections.
	Integrating over one complete revolution of the beam then yields,
	\begin{align}\label{sagnac subleading solution}
		\ S^{(1)}\Big\vert^{\lambda_1}_0=\int_0^{\lambda_1} d\lambda \frac{DS^{(1)}}{D\lambda}=\int_0^{\lambda_1}d\lambda \,q^\mu\partial_\mu S^{(1)}=\frac{1}{2}\int_0^{\lambda_1} d\lambda (\omega^2\delta g_{00}+\delta g_{ij}\,q^i\,q^j)+\omega\int dx^ig_{0i}.
	\end{align}
	Note that the beam traces a closed loop in space, so the last integral is in fact a line integral around the loop.
	For later convenience, we define the vector \(\cA_i \equiv -\tfrac{1}{2} g_{0i}\).
	In a Sagnac experiment the second beam propagates along the same loop but in the opposite direction, this effectively corresponds to
	\(q_i \rightarrow -q_i\) while \(\w \rightarrow \w\). The first integral on the right hand side of \eqref{sagnac subleading solution} is even under this inversion and therefore cancels out when computing the phase difference between counter-orbiting beams. We are therefore left with 
	\begin{empheq}[box=\othermathbox]{equation}\label{phase shift}
		\Delta S=\ S_1^{(1)}\Big\vert^{\lambda_1}_0- {S}_2^{(1)}\Big\vert^{{\lambda}_2}_0=-4\omega\oint_\cC \cA\,\,,
	\end{empheq}
	where in the integral is over the one-form $\cA=\cA_i dx^i$.
	In this derivation, one directly obtains the phase difference without referring to the time delay between the two beams~\cite{Ashtekar:1975wt,Frauendiener:2018gkw,Frauendiener:2018tut}.

	\subsection{Sagnac polarization effect}\label{section:Sagnac polarization}
	
	Going back to Figure~\ref{fig:holonomy}, assume that the initial polarization vectors are $f_{01},f_{02}$. For aesthetic reasons, we will denote a vector $p^a$ in the local frame by a ket $|p\rangle$ and the inner product by $\langle p_1|p_2\rangle=\eta_{ab}p_1^a p_2^b$. The final polarizations at point $B$ along $\gamma_1$ and $\gamma_2$ are given by \eqref{f evolution}, represented in matrix form
	\begin{align}
		|f_1\rangle &=U_{\gamma_1}|f_{01}\rangle,\qquad |f_2\rangle=U_{\gamma_2}|f_{02}\rangle\,.
	\end{align}  
	In the approximation discussed in \eqref{p1p2}, the polarization rotation appearing in \eqref{cross-correlation} is 
	\begin{empheq}[box=\othermathbox]{align}
		\langle p_2| p_1\rangle &=\langle f_{02}| U_{\Gamma}| f_{01}\rangle +O(r^{-2}), \qquad U_\Gamma=\mathcal{P}\exp \big[-\oint_{\Gamma} \omega\big]\,, \label{holonomy polarization rotation}
	\end{empheq}
	where $\Gamma\equiv \gamma_1\circ -\gamma_2$ is a closed spacetime loop going forward in time along $\gamma_1$ and returns back along $\gamma_2$, see Figure~\ref{fig:holonomy}. 
	An important property of the above result is that it is invariant under local Lorentz gauge transformations. This can be easily seen from the  gauge transformation of the holonomy \eqref{holonomy gauge transformation}.
	We conclude that local Lorentz transformations, \ie the choice of the tetrad will not affect the relative polarization rotation, and we are free to do the calculation in any frame we wish. 
	
	A notable feature of the polarization rotation compared to the phase shift is its independence of the frequency of light. This can have a positive and a negative side. On the one hand, the phase shift is enhanced by increasing light's frequency, while the polarization rotation is not. On the other hand, the frequency independence of the holonomy can be used as a signature to distinguish it from other effects, including noises in an actual experiment.

	\paragraph{Adiabatic limit.} In typical situations, the frequency of the gravitational field is much smaller than the frequency of oscillation of the light beam inside the loop. For example, for typical stellar mass black hole binaries, the frequency $\nu_{\text{GW}}\sim100$Hz, while the frequency of oscillation of the light beam inside a 1 meter loop is $\nu_{\text{loop}}\sim 10^8$Hz. As a result, the  temporal extent of the loop in Figure~\ref{fig:holonomy} shrinks to zero and the holonomy can be approximated by 
	\begin{align}\label{holonomy adiabatic}
		U_\Gamma&\approx U_{2\cC}=\mathcal{P}\exp \big[-2\oint_{\cC} \omega\big]\,,
	\end{align}
	where $\cC$ is the spatial loop along which the light beam is orbiting.


	\section{Fermi normal coordinates}\label{sec: FNC}
	
	Fermi normal coordinate (FNC) system provides a natural framework to study gravitational effects in the laboratory of a local observer. The metric in this coordinate system is written naturally in terms of the acceleration and rotation of the observer, as well as tidal effects from the Riemann curvature of spacetime~\cite{Manasse:1963zz,Misner:1973prb,Ni:1978zz}. 
	
	
	FNC is constructed as follows. Let $x^\mu(\tau)$ be the observer’s  worldline (which might be a geodesic or not) and $\tau$ its proper time. Its proper velocity and acceleration are defined as   $\hat{t}^\mu=\tfrac{dx^\mu}{d\tau}, \; a^\mu=\tfrac{d\hat{t}^\mu}{d\tau}=\hat{t}^\nu\nabla_\nu \hat{t}^\mu$. Construct an orthonormal frame adapted to the observer as $e_a{}^\mu=(\hat{t}^\mu , e_i{}^\mu)$. The evolution of the frame along the worldline is characterized as
	\begin{align}\label{frame transport}
		\frac{\mathrm{D} e^\mu_{a}}{\mathrm{D} \tau} =-(a^\mu \hat{t}_\nu-a_\nu\hat{t}^\mu+\hat{t}_\alpha\omega_\beta\epsilon^{\alpha\beta\mu}{}_\nu)e^\nu_{a}.
	\end{align}
	The antisymmetry of the terms inside the parentheses ensures orthonormality. The frame's evolution is therefore specified by  $a_i= e_i{}^\mu a_\mu$ and $\w_i=e_i{}^\mu \w_\mu$ The latter measures the failure of spatial basis $e_i$ from being Fermi-Walker transported along the worldline. 
	
	We define the Fermi normal coordinate $(T,X^i)$ of spacetime events as follows. If the event is on the worldline, we take $(T,X^i)=(\tau,0)$. For a given spacetime event $p$ off the worldline, there is a unique spacelike vector $X^i$ whose exponential map defines a spatial geodesic $x^\mu(\lambda)$ such that at $\lambda=0$ intersects the observer's worldline orthogonally at proper time $\tau$ and reaches the event $p$ at $\lambda=1$. The FNC of the point $p$ is thus defined as $(T=\tau, X^i)$. The spacetime metric in this coordinate system is given as an expansion in powers of $X^i$. To quadratic order $O(X^2)$\cite{Ni:1978zz}, 
	\begin{align}\label{fncmetric}
		ds^{2} &= - dT^{2} \Big[ 1 + 2 a_{j} X^{j} + \big( a_i a_j +\w_i \w_j - \w_k \w^k \de_{ij} + R_{0 i 0 j} \big)  X^{i} X^{j} \Big] \\
		&\quad + 2 dT dX^{i} \Big[ \epsilon_{i j k} \, \omega^{j} X^{k} - \frac{2}{3} R_{0 n i m} \, X^{n} X^{m} \Big]  + dX^{i} dX^{j} \Big[ \de_{i j} - \frac{1}{3} R_{i n j m} \, X^{n} X^{m} \Big]+ O( X^3 ).\nonumber
	\end{align}
	It is convenient for both physical interpretation and subsequent calculations to recast the local FNC metric in the language of \textit{gravito-electromagnetism} (GEM), see \eg  \cite{Mashhoon:2003ax}.  In this viewpoint, we write the metric \eqref{fncmetric} in terms of effective scalar and vector potentials $\Phi, \cA_i$  and the tidal field $h_{ij}$ as
	\begin{align}\label{GEM metric}
		ds^2=-(1-2\Phi)dT^2 -4 \cA_i dT dX^i +(\delta_{ij}+h_{ij})dX^i dX^j,
	\end{align}
	where 
	\begin{subequations}\label{GEM potentials}
		\begin{align}
			&\Phi(X)=-\frac{1}{2}\Big[  2 a_{j} X^{j} + \big( a_i a_j +\w_i \w_j - \w_k \w^k \de_{ij} + R_{0 i 0 j} \big)  X^{i} X^{j}\Big]+ O( X^3 ),\label{Phi def}\\
			&\cA_i(X)=-\frac{1}{2}\Big[ \epsilon_{i j k} \, \omega^{j} X^{k} - \frac{2}{3} R_{0 l i m} \, X^{l} X^{m} \Big]+ O( X^3 ),\label{Ai def}\\
			&h_{ij}(X)= - \frac{1}{3} R_{i l j m} \, X^{l} X^{m} + O( X^3 ).\label{hij def}
		\end{align}
	\end{subequations}

	\section{Sagnac effect from gravitational waves}\label{sec: Bondi}

	Asymptotically flat spacetimes near null infinity are conveniently described using
	Bondi coordinates \( (u,r,\theta^{A}) \), where \(u\) labels null hypersurfaces,
	\(r\) is a radial (luminosity) coordinate and \(\theta^{A}\) are angular coordinates on  the celestial sphere.
	In Bondi gauge the metric takes the form
	\begin{align}
		ds^2  =&  -(1-\frac{2m}{r})du^2 - 2 du\, dr + r^2 \Big(q_{AB}
		+\frac{C_{AB}}{r}\Big) d\theta^A d\theta^B + D^AC_{AB}\, du\, d\theta^B + \dots ,
	\end{align}
	where $q_{AB}$ is the metric of the round celestial sphere, and $D_A$ is the associated covariant derivative $D_A q_{BC}=0$. Bondi data includes the  Bondi mass aspect $m(u,\th^A)$ and the symmetric trace-free shear tensor $C_{AB}(u,\th^A)$. Moreover, $N_{AB} = \dot{C}_{AB}$  is the Bondi news tensor. Hereafter, overdot refers to derivative with respect to $u$.
	
	In the following, we consider two asymptotic observers in Bondi coordinates, I) a \textit{static} observer: an accelerated observer at constant spatial coordinates $(r,\th^A)$ , II) a \textit{freely falling} observer. We will use the results of Section~\ref{sec: FNC} to construct a local FNC associated to these observers, and use them to compute the Sagnac effect for each observer. 
	
	To construct the Fermi coordinates associated to any of the above observers, what we need is an explicit identification of the observer's worldline $x^\mu(\tau)$ and frame $e_a{}^\mu$ in terms of Bondi coordinates. From there, we can read off the relevant quantities appearing in the corresponding FNC metric \eqref{fncmetric},
	\begin{align}
		a_i=e_i^\mu a_\mu\,,\qquad \w_i=e_i^\mu \w_\mu\,,\qquad R_{abcd}=e_a{}^\mu e_b{}^\nu e_c{}^\alpha e_d{}^\beta R_{\mn\alpha\beta}\,.
	\end{align}
	The right hand sides of these equations are then computable in Bondi coordinates.
	
	\subsection{Hierarchy of scales.} \label{Hierarchy}
	The Fermi normal coordinates, discussed in Section~\ref{sec: FNC}, can involve several length scales. In the problem at hand, there are three length scales: the size of the laboratory $|X|$,  distance $r$ of the observer to the source of GW and the wavelength $\lgw$ of GWs. A necessary condition for the validity of the FNC expansion is that $\frac{|X|}{r}\ll 1$ and $\frac{|X|}{\lgw}\ll 1$. However, we have another assumption here: that $\lgw\ll r$. This hierarchy is well satisfied for astrophysical gravitational-wave sources: their distances are typically of the order of billions of light-years, whereas the relevant wavelengths are only a few thousand kilometers, corresponding to a frequency of order$\sim 100$Hz. 
	
	As we will see explicitly below, the leading behavior of the quantities appearing in the FNC metric are
	\begin{align}
		|a_i|\sim \frac{|N_{AB}|}{r}\sim |\w_i|\,,\qquad |R_{abcd}| \sim \frac{|\dot N_{AB}|}{r}.
	\end{align}
	Looking at \eqref{fncmetric}, these imply consistency relations 
	\begin{align}
		a_i X^i\sim \frac{|X|}{r}\ll1\,,\qquad \frac{R_{0i0j}X^iX^j}{a_iX^i}\sim \frac{|\dot N_{AB}|X}{|N_{AB}|}\sim \frac{X}{\lgw}\ll1  .
	\end{align}
	Moreover, 
	the hierarchy $\lgw \ll r$ further implies  that 
	\begin{align}
		\frac{(a_iX^i)^2}{R_{0i0j}X^iX^j}\sim \frac{|N_{AB}|^2}{|\dot N_{AB}|r}\sim |N_{AB}|\frac{\lgw}{r}\ll 1    .
	\end{align}
	Therefore, we can drop quadratic terms in the acceleration and rotation vectors in \eqref{fncmetric}. Terms that are not appearing in \eqref{fncmetric} are subleading with factors of $\frac{|X|}{r}$ or $\frac{|X|}{\lgw}$.
	
	\subsection*{Tetrad and spin connections in the weak field}
	
	In this subsection, we construct a tetrad and the corresponding spin connections associated to the FNC metric \eqref{GEM metric}. Given the hierarchy discussed in the previous subsection, we drop quadratic quantities in the acceleration, rotation and Riemann and denote the remainders simply as $O(\frac{1}{r\,\lgw})$ in what follows.
	In this approximation, there exists a natural tetrad adapted to GEM. The metric \eqref{GEM metric} can be written as $ds^2=\eta_{ab}\,e^a\otimes e^b$, where the co-frame 1-forms are
	\begin{align}\label{tetradgem}
		e^0&=(1-\Phi)dT+2\cA_i dX^i+O(\tfrac{1}{r\,\lgw})\,,\qquad e^i=dX^i+\frac12 h_{ij}dX^j+O(\tfrac{1}{r\,\lgw}).
	\end{align}
	At the origin of FNC $X^i=0$, the tetrad reduces to $e^0=dT,e^i=dX^i$ which is expected by the construction of FNC. 
	The spin connection $\w_{ab}$ one-forms associated to the above tetrad can be derived by imposing the torsionless condition $de^a+\w^a{}_b \wedge e^b=0$ as well as metricity, which in turn implies that $\w_{ab}=-\w_{ba}$. We relegate the derivation to appendix~\ref{spinderivation} and report the final result here
	\begin{subequations}\label{spin connections GEM}
		\begin{empheq}[box=\othermathbox]{align}
			\bw_{0i}&=-\cE_i dT+\big(\cF_{ij}-\frac12 \dot{h}_{ij}\big)dX^j+O(\tfrac{1}{r\,\lgw})\,,\label{w0i GEM}\\
			\bw_{ij}&=-\cF_{ij} dT+\pd_{[i} h_{j]k} dX^k+O(\tfrac{1}{r\,\lgw}) ,\label{wij GEM}
		\end{empheq}
	\end{subequations}
	where the GEM field strengths $\cE_i$ and $\cF_{ij} $ are defined as 
	\begin{align}
		\cE_i&\equiv-\pd_i\Phi-2\dot{\cA}_i\,,\qquad  \cF_{ij}\equiv \pd_i \cA_j-\pd_j \cA_i\,,
	\end{align}
	together with the tidal effect $\pd_{[i} h_{j]k}$, take the explicit form
	\begin{subequations}
		\begin{align}
			\cE_i&=a_i+\big( R_{0i0k}+\epsilon_{ijk}\dot{\omega}^j \big) X^k-\frac{2}{3}\dot{R}_{0jik}X^jX^k+O(\tfrac{1}{r\,\lgw}),\\
			\cF_{ij}&=-\eps_{ijk}\w^k-R_{0kij}X^k+O(\tfrac{1}{r\,\lgw})\,,\\
			\pd_{[i} h_{j]k}&=\frac{1}{2}R_{ijkl}X^l+O(\tfrac{1}{r\,\lgw})\,.\label{dihjk}
		\end{align}
	\end{subequations}

	\subsection{Static observer in Bondi--Sachs spacetime}\label{subsec:static_observer}
	
\begin{figure}
		\centering
		\includegraphics[width=0.5\linewidth,trim=0cm 1.5cm 1cm 2cm,
		clip]{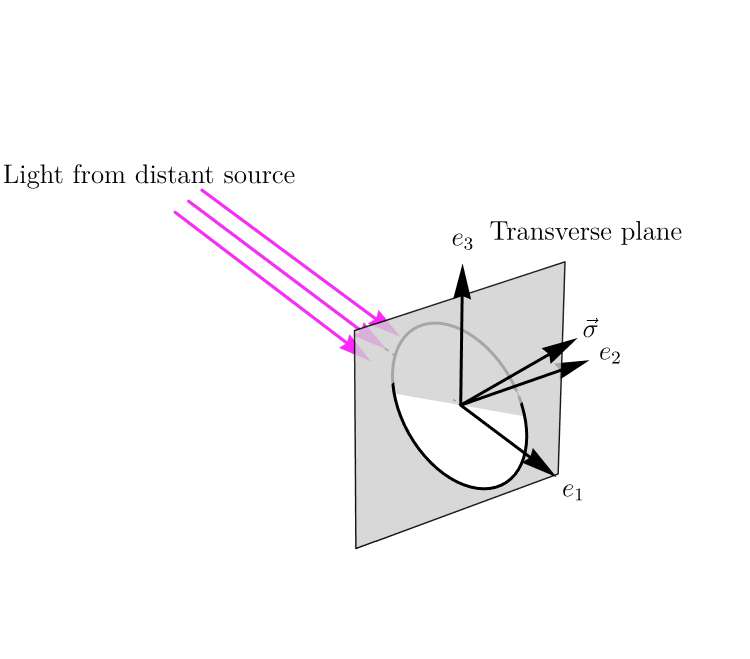}
			\captionsetup{width=.8\linewidth}
		\caption{Orthonormal frame of the observer, adapted to outgoing light rays. The Sagnac loop is centered on the observer and has an area vector $\boldsymbol{\sigma}$.}
		\label{fig:frame}
	\end{figure}

	For the static observer at $r=$const and $\th^A=$const, the normalized velocity is given by 
	\begin{equation}\label{staticobs}
		\hat{t}^\mu = \left(1+\frac{m}{r}\right)\delta^\mu{}_{u}+O(r^{-2}).
	\end{equation}
	We take the spatial basis vectors $e_i{}^\mu$ to consist of a unit radial vector $e_1$ which is derived from the spatial projection of the Bondi null vector $\partial_r$ and the remaining two vectors $e_2,e_3$, collectively denoted as $e_I$ to be orthonormal and tangent to the celestial sphere. The frame is depicted in Figure~\ref{fig:frame}.
	\begin{align}
		e_0&=\partial_u+O(r^{-1}),\qquad 
		e_1= -\partial_u+\partial _r+O(r^{-1}),\qquad 
		e_I=\frac{1}{r}E_I{}^A\pd_A+O(r^{-2})\,,
	\end{align}
	where $E_I{}^A E_J{}^B q_{AB}=\de_{IJ}$. For example, in polar coordinates  $e_{2}=\frac{1}{r}\partial_{\theta},\; e_{3}=\frac{1}{r\sin{\theta}}\partial_{\phi}$ up to subleading corrections. At $O(1/r)$ there is only one non-zero Riemann component in Bondi coordinates is $R_{0 A0 B} = -\frac{1}{2r} \ddot{C}_{AB} + O(r^{-2})$. Therefore, in the local frame, non-zero components of the Riemann tensor are
	\begin{align}\label{Riemannstatic}
		R_{0 I 0 J} =R_{1I1J} =-R_{1I0J} = -\frac{1}{2r}\dot N_{IJ} + O(r^{-2})\,,
	\end{align}
	and other components which are derivable from permutation symmetries of the Riemann tensor. 
	Notice that $I,J$ indices on fields living in the spacetime (like the Riemann tensor) are contraction with $e_I$, while fields living on the celestial sphere (like the news tensor) are obtained by contraction with $E_I$. 
	
	For the static observer \eqref{staticobs}, the proper acceleration and frame rotation to leading order are
	\begin{align}
		a_{1} &= -\frac{\dot m}{r} + O(r^{-2}), & a_{I}& = \frac{1}{2r}D^J N_{IJ} + O(r^{-2}),\\
		\w_{1}&=O(r^{-2})\,,&\w_{ I} &= -\frac{1}{2r}\eps_{IJ} D_K N^{JK} + O(r^{-2}),
	\end{align}
	where $\eps_{IJ}$ is the alternating symbol (with $\pm 1$ entries) and frame indices are obtained by contraction with frame fields on the sphere, e.g.,  \(D_I X_J= E_I{}^A E_J{}^B D_A X_B\). 
	
	Substituting $a_{i},\omega_{i}$ as well as the Riemann tensor components into \eqref{fncmetric}, we find the FNC metric for the asymptotic static observer
	\begin{align}
		\label{eq:static_fnc_metric}
		ds^2_{\text{static}} &=  -(dT)^2+(dX^1)^2+\de_{IJ}X^I X^J\nonumber\\ 
		&+\frac{1}{r}\Bigg\{
		dT^2\Big(2\dot{m}X^1- D^JN_{IJ}\,X^I + \frac{1}{2}\dot{N}_{IJ}\,X^IX^J\Big) \nonumber\\
		& + dT\Big(D^J N_{IJ}\big( X^1\,dX^I-X^I\,dX^1 \big)-\frac{2}{3}\dot{N}_{IJ}(X^IX^J dX^1-X^I X^1 dX^J)\Big)\nonumber\\
		& + \frac{\dot{N}_{IJ}}{6}\Big(X^IX^J (dX^1)^2 + (X^1)^2 dX^I dX^J - 2X^JX^1\,dX^I dX^1\Big)\Bigg\}+ O(r^{-2})\,.
	\end{align}
	All Bondi data entering the FNC metric are evaluated on the worldline of the observer. The gravitomagnetic potential \(A_i\) can be read from this metric as
	\begin{equation}
		\cA_i\,dX^i = \frac{1}{4r}D^IN_{IJ}\big(X^J\,dX^1 - X^1\,dX^J\big)+\frac{1}{6r}\dot{N}_{IJ}(X^IX^J dX^1-X^I X^1 dX^J)+ O(1/r^2).
	\end{equation}
	From this we can compute the two-form \(\cF=d \cA\), 
	\begin{align}\label{Fij static}
		\cF &= \frac{1}{2r}\Big(D^I N_{IJ}+\frac{2}{3}\dot{N}_{IJ}X^I\Big) dX^J\wedge dX^1.
	\end{align}
	Now consider a Sagnac loop whose `center' is positioned at the location of the observer, and its area vector $\boldsymbol{\sigma}$, defined as $\sigma^i=\frac12\eps^{ijk}\int X_j dX_k$ and depicted in Figure~\ref{fig:frame}. Now we make use of  the general Sagnac expression for the phase shift to write it as
	\begin{equation}
		\label{eq:Delta_psi_general}
		\Delta S = -4\omega\oint_\cC \cA=-4\omega\int \cF=-4\omega\, \boldsymbol{\cB} \cdot \boldsymbol{\sigma},
	\end{equation}
	where the gravitomagnetic field is $\cB_i\equiv\frac12\epsilon_{ijk}\cF^{jk}$ and the dot product in \eqref{eq:Delta_psi_general} is Euclidean. The gravitomagnetic vector field at the observer's worldline evaluates to 
	\begin{align}
	\boldsymbol{\cB}	=(\cB_1,\cB_J)=(0,\frac{1}{2r}\eps_{JK}D_IN^{IK}).
	\end{align}
	Therefore, the Sagnac phase shift for the static observer reads
	\begin{empheq}[box=\othermathbox]{align}
		\label{eq:Delta_psi_static_final}
		\Delta S_{\text{static}} =\frac{-2\omega }{r}\epsilon^{IJ} \sigma_I D^KN_{JK} + O\!\left(\frac{1}{r^2}\right).
	\end{empheq}
	where $\sigma_I$ is the projection of the area vector on the transverse plane.
	
	\paragraph{Polarization rotation.} Now we turn our attention to the polarization rotation. According to \eqref{holonomy adiabatic}, in the adiabatic limit, the polarization rotation is sensitive to the spatial part of the spin connection, given by \eqref{wij GEM} and \eqref{dihjk},
	\[
	\omega_{kij}=\frac{1}{2}R_{ijkl}X^l +O(\frac{1}{r^2}).
	\]
	The holonomy matrix to order reads
	\begin{equation}
		U_{ij}=\delta_{ij}-\oint\frac{1}{2}R_{ijkl}X^l dX^k=\delta_{ij}+\frac{1}{4}R_{ijkl}\epsilon^{klp}\sigma_p. 
	\end{equation}
	It is easy to check from \eqref{Riemannstatic} that the only non-trivial corrections are $U_{rA}$ and $U_{Ar}$.
	\begin{equation}
		U_{I1}=-U_{1I}=\frac{1}{4r}\dot{N}_{IJ}\sigma_K \epsilon^{JK}.
	\end{equation}
	
	The holonomy angle becomes
	\begin{equation}\label{polarization static}
		p_1 \cdot p_2=f_{01}\cdot f_{02}+\frac{1}{4r} \dot{N}_{IJ}\sigma_K \epsilon^{JK}(f_{01}^I f_{02}^1-f_{01}^1 f_{02}^I)+O\!\left(\frac{1}{r^2}\right).
	\end{equation}
	With the explanation given in Table~\ref{table:initial conditions}
	to maximize the effect of the polarization rotation we arrange the setup such that $f_{01}\cdot f_{02}=0$ and therefore we have
	\begin{empheq}[box=\othermathbox]{align}
		p_1 \cdot p_2=\frac{1}{4r} \dot{N}_{IJ}\sigma_K \epsilon^{JK}(f_{01}^I f_{02}^1-f_{01}^1 f_{02}^I)+O\!\left(\frac{1}{r^2}\right).\label{polarization rotation Bondi}
	\end{empheq}

	
	\subsection{Freely-falling observer in Bondi--Sachs spacetime}
	\label{subsec:freefalling_observer}
	
	A freely-falling observer follows a timelike geodesic \(\gamma(\tau)\) with \(a^i=0\) and carries a
	parallel-transported tetrad \(\omega^i=0\). Setting \(a_i=0,\ \omega^i=0\) in the FNC expansion
	removes linear inertial terms; the remaining nontrivial contributions are curvature/tidal terms
	\begin{align}
		\label{eq:freefall_fnc_metric}
		ds^2_{\text{free}} =\; & -(dT)^2+(dX^1)^2+\delta_{IJ}dX^IdX^J \nonumber\\
		& +  \frac{\dot{N}_{IJ}}{6r}\bigg[ 3X^IX^J\,dT^2+4dT(X^I X^1 dX^J-X^IX^J  dX^1)\nonumber\\
		& \qquad \quad + X^IX^J (dX^1)^2+ (X^1)^2dX^I dX^J - 2dX^I dX^1\,X^JX^1 \bigg] + O(r^{-2}).
	\end{align}
	Reading off the mixed components we obtain the exact leading-order statement
	\begin{equation}
		\label{eq:A_free_vanish}
		\cA_i^{\text{free}} dX^i= \frac{1}{6r}\dot{N}_{IJ}(X^IX^JdX^1-X^I X^1 dX^J) + O(r^{-2}),
	\end{equation}
	i.e. there is only a quadratic-in-\(X\) gravitomagnetic potential in the freely-falling FNC expansion and this means that the GEM field strength is linear in $X^i$ and therefore its contribution to the Sagnac phase is zero for a planar loop with its center is aligned with the origin where the observer is.
	From \eqref{eq:Delta_psi_general} and \eqref{eq:A_free_vanish} we have
	\begin{empheq}[box=\othermathbox]{align}
		\label{eq:Delta_psi_free_zero_exact}
		\Delta S_{\text{free}} = 2p_0\oint_\cC \cA_i\,dX^i 
		= O\!\left(\frac{1}{r^2}\right).
	\end{empheq}
	Thus the leading Sagnac term cancels exactly; the first nonzero contribution is subleading and
	curvature-driven.
	The polarization rotation, however, is exactly the same expression from the static observer. This is obvious from the fact that the spatial metrics $h_{ij}$ are identical in both cases and the fact that polarization rotation is solely read from $h_{ij}$.  We also make an assumption about the initial state of the polarizations similar to the static case, therefore we have
	\begin{empheq}[box=\othermathbox]{align}
		p_1 \cdot p_2=\frac{1}{4r} \dot{N}_{IJ}\sigma_K \epsilon^{JK}(f_{01}^I f_{02}^1-f_{01}^1 f_{02}^I)+O(\frac{1}{r^2}).
	\end{empheq}
	

	\subsection{Analysis}
	
	Let us now compare the magnitude of different effects. As explained in Table~\ref{table:initial conditions}, the Sagnac phase \eqref{eq:Delta_psi_static_final} and the holonomy angle \eqref{polarization static} are, in principle, independently measurable. Since both effects appear at the same asymptotic order \( O(1/r) \), it is therefore necessary to determine which of them provides the dominant contribution.
	
	The relative magnitude of the two effects is controlled by the ratio $\frac{|\dot{N}_{AB}|}{\w\, |N_{AB}|}$. Upon using  \( \dot{N}_{AB} \sim \omega_{\mathrm{\rm g}} N_{AB} \), we find that 
	\begin{equation}
		\frac{p_1\cdot p_2}{\Delta S}\Big\vert_{\text{static}}\approx \frac{\omega_{\rm g}}{\omega}= \frac{\lem}{\lgw}\approx10^{-12}.
	\end{equation}
	Therefore, the polarization rotation is typically much smaller than the Sagnac phase shift.
	
	Another remark concerns the polarization rotation. We observe that the polarization rotation for the static and freely falling observers coincide at leading order. This is partly related to the gauge covariance of the holonomy \eqref{holonomy gauge transformation}, which implies the invariance of \eqref{holonomy polarization rotation} under local Lorentz transformations. The Sagnac phase shift \eqref{phase shift} does not enjoy such invariance.

	\paragraph{Memory effects.} The result \eqref{eq:Delta_psi_static_final} is the Sagnac phase shift after one loop. However, the interference could happen after $N$ loops. While the one-loop interference can be used to measure the instantaneous value of the news tensor, the accumulated phase after many loops captures GW memory effects. To see this, let $T$ be the travel time of one loop.  The $N$-loop effect involves 
	\begin{align}
		\sum_{n=0}^N N_{JK}(n \,T)\approx \frac{1}{T}\int_0^{t=NT}N_{JK}(u)du
	\end{align}
	Therefore, the net phase shift is 
	\begin{align}
		\Delta S^{\rm N-loop}&=-\frac{1}{r}\,\frac{4\pi\sigma_I}{\lem \,T}\,\epsilon^{IJ} D^K\Delta C_{JK}
	\end{align}
	where $\Delta C_{JK}= C_{JK}\big\vert_{t=0}^{t=NT}$. For a circular loop of radius $R$, the factor $\frac{4\pi|\sigma_I|}{\lem \,T}=\frac{2\pi R}{\lem}\sim 10^6$ for the scales assumed in Figure~\ref{fig:hierarchy}. Therefore, the static (accelerated) Sagnac interferometer is sensitive to memory effects, with a magnifying factor $|X|/\lem$.
	On the other hand, the polarization rotation \eqref{polarization rotation Bondi} is proportional to $\dot{N}_{AB}$. The extra time derivative makes it less memory-sensitive, because $N_{AB}$ vanishes before and after the burst of GWs.
	
	Another remark about the polarization rotation is its independence on the frequency of the EM beam. This could be a signature that distinguishes this effect from the eikonal phase shift, as well as several other noises in such systems. This could be of importance for realistic applications, see~\cite{Fedderke:2024ncj,Mashhoon:2024wvp}.

	\section{Discussion}
	The framework implemented in this paper can be used to study various GW effects. While we focused on the effects of GWs on propagation of EM waves and in particular the Sagnac experiment, we believe that the setup can be very useful to study GW memory observables, such as those discussed in \cite{Pasterski:2015tva,Nichols:2017rqr,Seraj:2021rxd,Seraj:2022qyt,Godazgar:2022pbx}. 
	
	Concerning the effect of gravitational waves on a Sagnac interferometer at large distance, we showed that the Sagnac phase shift and polarization rotation, both appear at $O(1/r)$, while the latter effect is typically smaller by a factor $\frac{\lem}{\lgw}$. An important exception is the case of a freely falling laboratory, in which the Sagnac phase shift is zero and the leading effect is a polarization rotation. It is interesting to compare the polarization rotation effects discussed in this paper with those discussed recently in \cite{Fedderke:2024ncj,Mashhoon:2024wvp}.
	
	We showed that in the particular case, where the  Sagnac loop is tangent to the sphere (\ie lies in the transverse plane of Figure~\ref{fig:frame}), then the effect is suppressed to $O(1/r^2)$. However, we did not study such subleading effects carefully in this paper. One future direction is to include such subleading effects in our framework and provide a unified framework for subleading memory effects~\cite{Pasterski:2015tva,Seraj:2021rxd,Seraj:2022qyt,Seraj:2022qqj}. At subleading orders, there are several interesting effects showing up, including the gravitational spin-Hall effect~\cite{Shoom:2020zhr,Shoom:2022oer}.
	
	We showed that the Sagnac phase shift can be written as an integral over the gravitomagnetic gauge field. This suggests that the Another important improvement could be to realize the Sagnac phase, as the holonomy associated to a symmetry, possibly local Poincar\'e translations. Partial results along that line was provided in~\cite{Seraj:2022qqj}, while further work is needed to establish this connection.
	
	We provide the basis for extending our analysis to the propagation of light beams inside a medium with the application of optical fibers in Sagnac interferometers in mind. Further progress along that line would be interesting. Moreover, it is interesting to consider other closed loop interferometers, such as zero-area loops ($\infty$-shape), or those suggested in \cite{Frauendiener:2018gkw,Frauendiener:2018tut}, and how such detectors can be of practical use in real gravitational wave experiments.
	
	\subsection*{Acknowledgements} The authors are grateful to Sajad Aghapour for early contribution in this research. AS would like to thank the Institut des Hautes Études Scientifiques (IHES) for an inspiring visit during which part of this research was carried out, and to T. Damour for related fruitful discussions.

	\appendix

	\section{Polarization transport in optical fibers}\label{appendix:fiber}
	
	In this paper, we focused on Sagnac configurations in which light propagates in vacuum and is guided by mirrors. An important question is how these results change when fiber optics are used instead. We refer the reader to recent relevant works~\cite{Bliokh:2009ek,Mieling:2024jvk}.  
	Practical optical fibers can have a wide variety of refractive–index profiles, we do not analyze specific fiber geometries here; instead, we refer the reader to the literature where such cases are treated in detail.  For simplicity, we assume the fiber's refractive-index profile is piecewise constant (e.g. single mode fiber). We solve Maxwell's equations within each uniform segment and do not consider the matching conditions at interfaces between segments with different refractive indices. This presents no difficulty, the differential equations governing electromagnetic wave propagation inside the fiber are independent of the boundary conditions. The latter become relevant only when one seeks particular solutions of these equations.
	
	In this appendix, we denote the $4d$ spacetime's metric and its covariant derivative as  ${\tlg}_{\mu\nu}$ and $\tilde{\nabla}_\mu$, and reserve $g_\mn, \cd_\mu$ for the \textit{optical metric} (defined below) and its associated derivative \cite{Gordon:1923qva},
	\begin{align}
		g_{\mu\nu}\equiv \tilde{g}_{\mu\nu}+(1-\frac{1}{n^2}) u^{\alpha}u^{\beta}\tilde{g}_{\nu\beta}\tilde{g}_{\mu\alpha}.
	\end{align}
	The inverse optical metric can be easily worked out,
	\begin{align}
		g^{\mu\nu}= \tilde{g}^{\mu\nu}-(n^2-1) u^{\mu}u^{\nu}.
	\end{align}
	
	Let's consider the Maxwell equations in an isotropic medium with constant refractive index \(n\) as measured by a congruence of observers \(O\).  This congruence is described by a unit timelike vector field \(u^\mu\), satisfying \(u^\mu u^\nu \tilde g_{\mu\nu} = -1\). If the observers are at rest with respect to the medium, then we simply have
	\[
	u^\mu = \frac{1}{\sqrt{-\tilde{g}_{00}}} (1,0,0,0).
	\]
	The Maxwell equations in this medium can be written in the following form~\cite{Gordon:1923qva},
	\begin{align}\label{tildeg}
		\tilde{\nabla}_\mu(g^{\mu\alpha}g^{\nu\beta}F_{\alpha\beta})=0, \qquad \tilde{\nabla}_\mu(g^{\mu\nu}A_\nu)=0,
	\end{align}
	Note that in \eqref{tildeg}, the covariant derivative is with respect to the spacetime metric. If we write the equations purely in terms of the spacetime metric and its covariant derivative, we find
	\begin{subequations}\label{spacetimeeq}
		\begin{align}
			&\tilde{\nabla}_\mu(\tilde{g}^{\mu\alpha}\tilde{g}^{\nu\beta}F_{\alpha \beta})=(n^2-1)\Big(\tilde{\nabla}_\mu(\tilde{g}^{\nu\beta}u^\mu u^\alpha F_{\alpha\beta})+\tilde{\nabla}_\mu(\tilde{g}^{\mu\alpha}u^\nu u^\beta F_{\alpha\beta})\Big),\\
			&\tilde{\nabla}_\mu(\tilde{g}^{\mu\nu}A_\nu)=(n^2-1)\tilde{\nabla}_\mu(u^\mu u^\nu A_\nu).
		\end{align}
	\end{subequations}
	On the other hand, we may try to write the equations purely in terms of the optical metric and its associated derivative.
	
	Defining $F^\mn\equiv g^{\mu\alpha}g^{\nu\beta}F_{\alpha\beta}$ and $A^\mu\equiv g^\mn A_\nu$, we note that 
	\begin{align}
		&\nabla_\mu (F^{\mu\nu})=\frac{1}{\sqrt{-g}}\partial_\mu(\sqrt{-g}F^{\mu\nu}),\quad \nabla_\mu A^\mu=\frac{1}{\sqrt{-g}}\partial_\mu(\sqrt{-g}A^{\mu}),\quad \det(g)=n^2 \det(\tilde{g}).
	\end{align}
	Since $n$ is constant, the above equations imply that $\cd_\mu F^\mn=\tilde{\cd}_\mu F^\mn$ and similarly $\nabla_\mu A^\mu=\tilde\nabla_\mu A^\mu$, and therefore \eqref{tildeg} implies Maxwell's equations inside the medium, purely written in terms of the optical metric and its associated derivative
	\begin{align}
		\nabla_\mu F^{\mu\nu}=0,\qquad \nabla_\mu A^\mu=0.
	\end{align}
	These equations should be compared with their more complicated spacetime counterparts in \eqref{spacetimeeq}. We find that Maxwell’s equations inside the fiber, when written with respect to the optical metric, take exactly the same form as Maxwell’s equations in vacuum. This observation allows us to extend all results obtained in vacuum to the case in which light propagates inside a fiber. In particular, by adopting the eikonal ansatz for the gauge field, one recovers precisely \eqref{eikonal equation} and \eqref{parallel transport}. For completeness and comparison, we also present their counterparts in spacetime, which look much more complicated, 
	\begin{subequations}
		\begin{align}
			&\tilde{g}^{\mu\nu}k_\mu k_\nu=(n^2-1)(k_\mu u^\mu)^2,\\
			&k_\beta (\tilde{g}^{\alpha\beta}-(n^2-1) u^\alpha u^\beta)\tilde{\nabla}_\beta \Big((\tilde{g}^{\nu\gamma}-(n^2-1) u^\nu u^\gamma)f_\gamma\Big)=-(n^2-1)f_\alpha k_\beta u^\nu \tilde{\nabla}^\beta u^\alpha.
		\end{align}
	\end{subequations}
	
	\section{Derivation of the spin connections}\label{spinderivation}
	To derive the spin connections corresponding to the GEM tetrad \eqref{tetradgem} in the weak field limit, we proceed as follows. We use the torsionless condition
	\begin{align}\label{torsionless}
		de^a+\w^a{}_b \wedge e^b&=0,
	\end{align}
	Taking the exterior derivative of \eqref{tetradgem} we have
	\begin{align}\label{spingem}
		de^0=\cE_i ~dX^i\wedge dT +\cF_{ij}~dX^i\wedge dX^j\,,\\
		de^i=\frac12 (\dot{h}_{ij} dT+\pd_k h_{ij}dX^k)\wedge dX^j\,.\label{dei}
	\end{align}
	On the other hand, it is convenient to exploit the algebraic property $\partial_{(i} h_{jk)} = 0$ 
	(which can be checked from \eqref{hij def}) to rewrite the exterior derivative (\ref{dei}) as
	\begin{align*}
		de^i=\frac12 (\dot{h}_{ij} dT-(\pd_i h_{jk}+\pd_j h_{ki})dX^i)\wedge dX^j.
	\end{align*}
	Substituting \(de^0\) and the above expression for \(de^i\) into the torsionless condition \eqref{torsionless} yields the component form of the spin connection,
	\begin{align}
		\w_{0i}&=\cE_i dT+\big(\cF_{ij}+\alpha_{ij}\big)dX^j\,,\\
		\w_{ij}&=-\big(\cF_{ij}+\alpha_{ij}+\frac12 \dot{h}_{ij}\big)dT+\frac12(\pd_i h_{jk}+\pd_j h_{ki}+\beta_{ijk})dX^k ,
	\end{align}
	where \(\alpha_{ij}\) and \(\beta_{ijk}\) are arbitrary so far. We now fix this remaining freedom by imposing the metricity condition which implies that $\omega_{ij}=-\omega_{ji}$.  This choice determines \(\alpha_{ij}\) and \(\beta_{ijk}\) as
	\begin{align}
		\alpha_{ij}=-\frac12 \dot{h}_{ij}\,,\qquad \beta_{ijk}=-\pd_i h_{jk},
	\end{align}
	and hence the spin connections take the simplified form 
	\begin{align}
		\bw_{0i}&=-\cE_i dT+\big(\cF_{ij}-\frac12 \dot{h}_{ij}\big)dX^j\,,\\
		\bw_{ij}&=-\cF_{ij} dT+\pd_{[i} h_{j]k} dX^k .
	\end{align}

	\addcontentsline{toc}{section}{References}

	\bibliography{references}

@book{Schneider:1992bmb,
    author = {Schneider, Peter and Ehlers, J{\"u}rgen and Falco, Emilio E.},
    title = "{Gravitational Lenses}",
    doi = "10.1007/978-3-662-03758-4",
    isbn = "978-3-540-66506-9, 978-3-662-03758-4",
    publisher = "Springer",
    series = "Astronomy and Astrophysics Library",
    year = "1992"
}

@article{Kulish:1970ut,
    author = "Kulish, P. P. and Faddeev, L. D.",
    title = "{Asymptotic conditions and infrared divergences in quantum electrodynamics}",
    reportNumber = "D70-07927",
    doi = "10.1007/BF01066485",
    journal = "Theor. Math. Phys.",
    volume = "4",
    pages = "745",
    year = "1970"
}

@article{Blanchet:1992br,
      author         = "Blanchet, Luc and Damour, Thibault",
      title          = "{Hereditary effects in gravitational radiation}",
      journal        = "Phys. Rev.",
      volume         = "D46",
      year           = "1992",
      pages          = "4304-4319",
      doi            = "10.1103/PhysRevD.46.4304",
      reportNumber   = "IHES-P-92-42",
      SLACcitation   = "%%CITATION = PHRVA,D46,4304;%%"
}

@article{Flanagan:2018yzh,
      author         = "Flanagan, \'Eanna \'E. and Grant, Alexander M. and Harte,
                        Abraham I. and Nichols, David A.",
      title          = "{Persistent gravitational wave observables: general
                        framework}",
      journal        = "Phys. Rev.",
      volume         = "D99",
      year           = "2019",
      number         = "8",
      pages          = "084044",
      doi            = "10.1103/PhysRevD.99.084044",
      eprint         = "1901.00021",
      archivePrefix  = "arXiv",
      primaryClass   = "gr-qc",
      SLACcitation   = "%%CITATION = ARXIV:1901.00021;%%"
}

@article{Pasterski:2015tva,
      author         = "Pasterski, Sabrina and Strominger, Andrew and Zhiboedov,
                        Alexander",
      title          = "{New Gravitational Memories}",
      journal        = "JHEP",
      volume         = "12",
      year           = "2016",
      pages          = "053",
      doi            = "10.1007/JHEP12(2016)053",
      eprint         = "1502.06120",
      archivePrefix  = "arXiv",
      primaryClass   = "hep-th",
      SLACcitation   = "%%CITATION = ARXIV:1502.06120;%%"
}

@article{Strominger:2017zoo,
      author         = "Strominger, Andrew",
      title          = "{Lectures on the Infrared Structure of Gravity and Gauge
                        Theory}",
      year           = "2017",
      eprint         = "1703.05448",
      archivePrefix  = "arXiv",
      primaryClass   = "hep-th",
      SLACcitation   = "%%CITATION = ARXIV:1703.05448;%%"
}

@article{Nichols:2017rqr,
      author         = "Nichols, David A.",
      title          = "{Spin memory effect for compact binaries in the
                        post-Newtonian approximation}",
      journal        = "Phys. Rev.",
      volume         = "D95",
      year           = "2017",
      number         = "8",
      pages          = "084048",
      doi            = "10.1103/PhysRevD.95.084048",
      eprint         = "1702.03300",
      archivePrefix  = "arXiv",
      primaryClass   = "gr-qc",
      SLACcitation   = "%%CITATION = ARXIV:1702.03300;%%"
}

@article{Seraj:2022qyt,
	author = "Seraj, Ali and Oblak, Blagoje",
	title = "{Precession Caused by Gravitational Waves}",
	eprint = "2203.16216",
	archivePrefix = "arXiv",
	primaryClass = "gr-qc",
	doi = "10.1103/PhysRevLett.129.061101",
	journal = "Phys. Rev. Lett.",
	volume = "129",
	number = "6",
	pages = "061101",
	year = "2022"
}

@article{Seraj:2021rxd,
	author = "Seraj, Ali and Oblak, Blagoje",
	title = "{Gyroscopic gravitational memory}",
	eprint = "2112.04535",
	archivePrefix = "arXiv",
	primaryClass = "hep-th",
	doi = "10.1007/JHEP11(2023)057",
	journal = "JHEP",
	volume = "11",
	pages = "057",
	year = "2023"
}

@article{Frauendiener:2018tut,
	author = {Frauendiener, J{\"o}rg},
	title = "{Gravitational waves and the Sagnac effect}",
	eprint = "1808.08653",
	archivePrefix = "arXiv",
	primaryClass = "gr-qc",
	doi = "10.1088/1361-6382/ab574c",
	journal = "Class. Quant. Grav.",
	volume = "37",
	number = "5",
	pages = "05LT01",
	year = "2020"
}

@article{Frauendiener:2018gkw,
	author = "Frauendiener, Joerg",
	title = "{Notes on the Sagnac effect in General Relativity}",
	eprint = "1808.07914",
	archivePrefix = "arXiv",
	primaryClass = "gr-qc",
	doi = "10.1007/s10714-018-2470-5",
	journal = "Gen. Rel. Grav.",
	volume = "50",
	number = "11",
	pages = "147",
	year = "2018"
}

@book{Henneaux:1994lbw,
	author = "Henneaux, Marc and Teitelboim, Claudio",
	title = "{Quantization of Gauge Systems}",
	isbn = "978-0-691-03769-1, 978-0-691-21386-6",
	publisher = "Princeton University Press",
	month = "8",
	year = "1994"
}

@article{Ni:1978zz,
    author = "Ni, Wei-Tou and Zimmermann, Mark",
    title = "{Inertial and gravitational effects in the proper reference frame of an accelerated, rotating observer}",
    doi = "10.1103/PhysRevD.17.1473",
    journal = "Phys. Rev. D",
    volume = "17",
    pages = "1473--1476",
    year = "1978"
}

@article{Manasse:1963zz,
    author = "Manasse, F. K. and Misner, C. W.",
    title = "{Fermi Normal Coordinates and Some Basic Concepts in Differential Geometry}",
    doi = "10.1063/1.1724316",
    journal = "J. Math. Phys.",
    volume = "4",
    pages = "735--745",
    year = "1963"
}

@book{Misner:1973prb,
    author = "Misner, Charles W. and Thorne, K. S. and Wheeler, J. A.",
    title = "{Gravitation}",
    isbn = "978-0-7167-0344-0, 978-0-691-17779-3",
    publisher = "W. H. Freeman",
    address = "San Francisco",
    year = "1973"
}

@article{Damour:2007nc,
    author = "Damour, Thibault and Jaranowski, Piotr and Schaefer, Gerhard",
    title = "{Hamiltonian of two spinning compact bodies with next-to-leading order gravitational spin-orbit coupling}",
    eprint = "0711.1048",
    archivePrefix = "arXiv",
    primaryClass = "gr-qc",
    doi = "10.1103/PhysRevD.77.064032",
    journal = "Phys. Rev. D",
    volume = "77",
    pages = "064032",
    year = "2008"
}

@article{Bini:2017xzy,
    author = "Bini, Donato and Damour, Thibault",
    title = "{Gravitational spin-orbit coupling in binary systems, post-Minkowskian approximation and effective one-body theory}",
    eprint = "1709.00590",
    archivePrefix = "arXiv",
    primaryClass = "gr-qc",
    doi = "10.1103/PhysRevD.96.104038",
    journal = "Phys. Rev. D",
    volume = "96",
    number = "10",
    pages = "104038",
    year = "2017"
}

@article{Ashtekar:1975wt,
    author = "Ashtekar, Abhay and Magnon, Anne",
    title = "{The Sagnac effect in general relativity}",
    doi = "10.1063/1.522521",
    journal = "J. Math. Phys.",
    volume = "16",
    pages = "341--344",
    year = "1975"
}

@book{Hecht2017,
  author    = {Eugene Hecht},
  title     = {Optics},
  edition   = {5},
  publisher = {Pearson},
  year      = {2017}
}

@article{Bliokh:2009ek,
    author = "Bliokh, Konstantin Y.",
    title = "{Geometrodynamics of polarized light: Berry phase and spin Hall effect in a gradient-index medium}",
    eprint = "0903.1910",
    archivePrefix = "arXiv",
    primaryClass = "physics.optics",
    doi = "10.1088/1464-4258/11/9/094009",
    journal = "J. Opt.",
    volume = "11",
    pages = "094009",
    year = "2009"
}

@article{Mieling:2024jvk,
    author = "Mieling, Thomas B. and Hudelist, Mario",
    title = "{Fiber optics in curved space-times}",
    eprint = "2410.23048",
    archivePrefix = "arXiv",
    primaryClass = "gr-qc",
    doi = "10.1103/PhysRevResearch.7.013162",
    journal = "Phys. Rev. Res.",
    volume = "7",
    number = "1",
    pages = "013162",
    year = "2025"
}

@article{Mashhoon:2003ax,
    author = "Mashhoon, Bahram",
    title = "{Gravitoelectromagnetism: A Brief review}",
    eprint = "gr-qc/0311030",
    archivePrefix = "arXiv",
    month = "11",
    year = "2003"
}

@article{Ruggiero:2015gha,
    author = "Ruggiero, Matteo Luca",
    title = "{Sagnac Effect, Ring Lasers and Terrestrial Tests of Gravity}",
    eprint = "1505.01268",
    archivePrefix = "arXiv",
    primaryClass = "gr-qc",
    doi = "10.3390/galaxies3020084",
    journal = "Galaxies",
    volume = "3",
    number = "2",
    pages = "84--102",
    year = "2015"
}

@article{Sun:1996bj,
    author = "Sun, K. X. and Fejer, M. M. and Gustafson, E. and Byer, R. L.",
    title = "{Sagnac interferometer for gravitational wave detection}",
    doi = "10.1103/PhysRevLett.76.3053",
    journal = "Phys. Rev. Lett.",
    volume = "76",
    pages = "3053--3056",
    year = "1996"
}

@article{Eberle:2010zz,
    author = "Eberle, Tobias and Steinlechner, Sebastian and Bauchrowitz, Joran and Handchen, Vitus and Vahlbruch, Henning and Mehmet, Moritz and Muller-Ebhardt, Helge and Schnabel, Roman",
    title = "{Quantum Enhancement of the Zero-Area Sagnac Interferometer Topology for Gravitational Wave Detection}",
    eprint = "1007.0574",
    archivePrefix = "arXiv",
    primaryClass = "quant-ph",
    doi = "10.1103/PhysRevLett.104.251102",
    journal = "Phys. Rev. Lett.",
    volume = "104",
    pages = "251102",
    year = "2010"
}

@article{Chen:2002mf,
    author = "Chen, Yan-bei",
    title = "{Sagnac interferometer as a speed meter type, quantum nondemolition gravitational wave detector}",
    eprint = "gr-qc/0208051",
    archivePrefix = "arXiv",
    doi = "10.1103/PhysRevD.67.122004",
    journal = "Phys. Rev. D",
    volume = "67",
    pages = "122004",
    year = "2003"
}

@article{Sagnac1913a,
  author = {Sagnac, G.},
  title = {L'{e}ther lumineux d{\'e}montr{\'e} par l'effet du vent relatif d'{\'e}ther dans un interf{\'e}rom{\`e}tre en rotation uniforme},
  journal = {C. R. Acad. Sci.},
  volume = {157},
  pages = {708--710},
  year = {1913}
}

@article{Sagnac1913b,
  author = {Sagnac, G.},
  title = {Sur la preuve de la r{\'e}alit{\'e} de l'{\'e}ther lumineux par l'exp{\'e}rience de l'interf{\'e}rographe tournant},
  journal = {C. R. Acad. Sci.},
  volume = {157},
  pages = {1410--1413},
  year = {1913}
}

@article{Post1967,
  author = {Post, E. J.},
  title = {Sagnac Effect},
  journal = {Rev. Mod. Phys.},
  volume = {39},
  pages = {475--493},
  year = {1967}
}

@article{Stedman1997,
  author = {Stedman, G. E.},
  title = {Ring-laser tests of fundamental physics and geophysics},
  journal = {Rep. Prog. Phys.},
  volume = {60},
  pages = {615--688},
  year = {1997}
}

@article{Anderson1994,
  author = {Anderson, R. and Bilger, H. R. and Stedman, G. E.},
  title = {``Sagnac effect'': A century of Earth-rotated interferometers},
  journal = {Am. J. Phys.},
  volume = {62},
  pages = {975--985},
  year = {1994}
}

@article{RizziRuggiero2003a,
  author = {Rizzi, G. and Ruggiero, M. L.},
  title = {The relativistic Sagnac effect: two derivations},
  journal = {Gen. Rel. Grav.},
  volume = {35},
  pages = {1745--1757},
  year = {2003}
}

@article{RizziRuggiero2003b,
  author = {Rizzi, G. and Ruggiero, M. L.},
  title = {Relativistic analysis of the Sagnac effect},
  journal = {Found. Phys.},
  volume = {34},
  pages = {1835--1847},
  year = {2004}
}

@article{AshtekarMagnon1975,
  author = {Ashtekar, A. and Magnon, A.},
  title = {The Sagnac effect in general relativity},
  journal = {J. Math. Phys.},
  volume = {16},
  pages = {341--344},
  year = {1975}
}

@article{Anandan1981,
  author = {Anandan, J.},
  title = {Sagnac effect in relativistic and nonrelativistic physics},
  journal = {Phys. Rev. D},
  volume = {24},
  pages = {338--346},
  year = {1981}
}

@article{Sivasubramanian2003,
  author = {Sivasubramanian, S. and Widom, A. and Srivastava, Y. N.},
  title = {Gravitational waves and the Sagnac effect},
  journal = {Phys. Rev. D},
  volume = {67},
  pages = {084023},
  year = {2003}
}

@article{ZeldovichPolnarev1974,
  author = {Zel'dovich, Ya. B. and Polnarev, A. G.},
  title = {Radiation of gravitational waves by a cluster of superdense stars},
  journal = {Sov. Astron.},
  volume = {18},
  pages = {17--23},
  year = {1974}
}

@article{Christodoulou1991,
  author = {Christodoulou, D.},
  title = {Nonlinear nature of gravitation and gravitational-wave experiments},
  journal = {Phys. Rev. Lett.},
  volume = {67},
  pages = {1486--1489},
  year = {1991}
}

@article{Thorne1992,
  author = {Thorne, K. S.},
  title = {Gravitational-wave bursts with memory: The Christodoulou effect},
  journal = {Phys. Rev. D},
  volume = {45},
  pages = {520--524},
  year = {1992}
}

@article{WisemanWill1991,
  author = {Wiseman, A. G. and Will, C. M.},
  title = {Christodoulou's nonlinear gravitational-wave memory},
  journal = {Phys. Rev. D},
  volume = {44},
  pages = {R2945--R2949},
  year = {1991}
}

@article{Favata2009a,
  author = {Favata, M.},
  title = {Post-Newtonian corrections to the gravitational-wave memory for compact binaries},
  journal = {Phys. Rev. D},
  volume = {80},
  pages = {024002},
  year = {2009}
}

@article{Favata2009b,
  author = {Favata, M.},
  title = {The gravitational-wave memory effect},
  journal = {Astrophys. J. Lett.},
  volume = {696},
  pages = {L159--L162},
  year = {2009}
}

@article{Favata2010,
  author = {Favata, M.},
  title = {The gravitational-wave memory effect},
  journal = {Class. Quantum Grav.},
  volume = {27},
  pages = {084036},
  year = {2010}
}

@article{Faye:2024utu,
	author = "Faye, Guillaume and Seraj, Ali",
	title = "{Gyroscopic gravitational memory from quasi-circular binary systems}",
	eprint = "2409.02624",
	archivePrefix = "arXiv",
	primaryClass = "gr-qc",
	doi = "10.1088/1361-6382/ada339",
	journal = "Class. Quant. Grav.",
	volume = "42",
	number = "3",
	pages = "035005",
	year = "2025"
}

@article{Plebanski1960,
  author = {Plebanski, J.},
  title = {Electromagnetic waves in gravitational fields},
  journal = {Phys. Rev.},
  volume = {118},
  pages = {1396--1408},
  year = {1960}
}

@article{Ishihara1988,
  author = {Ishihara, H. and Takahashi, M. and Tomimatsu, A.},
  title = {Gravitational Faraday rotation induced by a Kerr black hole},
  journal = {Phys. Rev. D},
  volume = {38},
  pages = {472--477},
  year = {1988}
}

@article{Sereno2004,
  author = {Sereno, M.},
  title = {Gravitational Faraday rotation in weak gravitational fields},
  journal = {Phys. Rev. D},
  volume = {69},
  pages = {023002},
  year = {2004}
}

@article{Sereno2005,
  author = {Sereno, M.},
  title = {Rotation of polarization by a gravitational lens},
  journal = {Mon. Not. R. Astron. Soc.},
  volume = {356},
  pages = {381--386},
  year = {2005}
}

@article{Faraoni2008,
  author = {Faraoni, V.},
  title = {The rotation of polarization by gravitational waves},
  journal = {New Astron.},
  volume = {13},
  pages = {178--181},
  year = {2008}
}

@article{Shoom:2022oer,
	author = "Shoom, Andrey A.",
	title = "{Faraday effect of light caused by plane gravitational wave}",
	eprint = "2206.08867",
	archivePrefix = "arXiv",
	primaryClass = "gr-qc",
	doi = "10.1007/s10714-024-03283-z",
	journal = "Gen. Rel. Grav.",
	volume = "56",
	number = "8",
	pages = "97",
	year = "2024"
}

@article{Shoom:2020zhr,
	author = "Shoom, Andrey A.",
	title = "{Gravitational Faraday and spin-Hall effects of light}",
	eprint = "2006.10077",
	archivePrefix = "arXiv",
	primaryClass = "gr-qc",
	doi = "10.1103/PhysRevD.104.084007",
	journal = "Phys. Rev. D",
	volume = "104",
	number = "8",
	pages = "084007",
	year = "2021"
}

@article{Lyutikov2017,
  author = {Lyutikov, M.},
  title = {Polarization rotation in relativistic gravitational lenses},
  journal = {Phys. Rev. D},
  volume = {95},
  pages = {084019},
  year = {2017}
}

@article{cohen1993standard,
  title={Standard clocks, interferometry, and gravitomagnetism},
  author={Cohen, Jeffrey M and Mashhoon, Bahram},
  journal={Physics Letters A},
  volume={181},
  number={5},
  pages={353--358},
  year={1993},
  publisher={Elsevier}
}

@article{Tinto:2014lxa,
    author = "Tinto, Massimo and Dhurandhar, Sanjeev V.",
    title = "{Time-Delay Interferometry}",
    doi = "10.12942/lrr-2014-6",
    journal = "Living Rev. Rel.",
    volume = "17",
    number = "1",
    pages = "6",
    year = "2014"
}

@article{allan1985around,
  title={Around-the-world relativistic Sagnac experiment},
  author={Allan, David W and Weiss, Marc A and Ashby, Neil},
  journal={Science},
  volume={228},
  number={4695},
  pages={69--70},
  year={1985},
  publisher={American Association for the Advancement of Science}
}

@article{Ruggiero:2023ker,
    author = "Ruggiero, Matteo Luca and Astesiano, Davide",
    title = "{A tale of analogies: a review on gravitomagnetic effects, rotating sources, observers and all that}",
    eprint = "2304.02167",
    archivePrefix = "arXiv",
    primaryClass = "gr-qc",
    doi = "10.1088/2399-6528/ad08cf",
    journal = "J. Phys. Comm.",
    volume = "7",
    number = "11",
    pages = "112001",
    year = "2023"
}

@article{Seraj:2022qqj,
    author = "Seraj, Ali and Neogi, Turmoli",
    title = "{Memory effects from holonomies}",
    eprint = "2206.14110",
    archivePrefix = "arXiv",
    primaryClass = "hep-th",
    doi = "10.1103/PhysRevD.107.104034",
    journal = "Phys. Rev. D",
    volume = "107",
    number = "10",
    pages = "104034",
    year = "2023"
}

@article{Gordon:1923qva,
    author = "Gordon, W.",
    title = {{Zur Lichtfortpflanzung nach der Relativit{\"a}tstheorie}},
    doi = "10.1002/andp.19233772202",
    journal = "Annalen Phys.",
    volume = "377",
    number = "22",
    pages = "421--456",
    year = "1923"
}

@book{Hinch_1991, place={Cambridge}, series={Cambridge Texts in Applied Mathematics}, title={Perturbation Methods}, publisher={Cambridge University Press}, author={Hinch, E. J.}, year={1991}, collection={Cambridge Texts in Applied Mathematics}}

@article{Godazgar:2022pbx,
    author = "Godazgar, Mahdi and Macaulay, George and Long, George and Seraj, Ali",
    title = "{Gravitational memory effects and higher derivative actions}",
    eprint = "2206.12339",
    archivePrefix = "arXiv",
    primaryClass = "gr-qc",
    doi = "10.1007/JHEP09(2022)150",
    journal = "JHEP",
    volume = "09",
    pages = "150",
    year = "2022"
}

@article{Fedderke:2024ncj,
    author = "Fedderke, Michael A. and Harnik, Roni and Kaplan, David E. and Posen, Sam and Rajendran, Surjeet and Serra, Francesco and Yakovlev, Vyacheslav P.",
    title = "{Precision gyroscope from the helicity of light}",
    eprint = "2406.16178",
    archivePrefix = "arXiv",
    primaryClass = "physics.optics",
    reportNumber = "FERMILAB-PUB-24-0325-SQMS-TD",
    doi = "10.1103/PhysRevA.111.043502",
    journal = "Phys. Rev. A",
    volume = "111",
    number = "4",
    pages = "043502",
    year = "2025"
}

@article{Mashhoon:2024wvp,
    author = "Mashhoon, Bahram and Obukhov, Yuri N.",
    title = "{Spin-of-light gyroscope and the spin-rotation coupling}",
    eprint = "2408.07799",
    archivePrefix = "arXiv",
    primaryClass = "quant-ph",
    doi = "10.1103/PhysRevD.110.104015",
    journal = "Phys. Rev. D",
    volume = "110",
    number = "10",
    pages = "104015",
    year = "2024"
}

\end{document}